\documentclass[3p,times]{elsarticle}

\usepackage{graphicx}
\usepackage{amsmath}
\usepackage{mathrsfs}
\usepackage{mathtools}
\usepackage{booktabs}
\usepackage{comment}
\usepackage[colorlinks,urlcolor=blue,linkcolor=blue,citecolor=blue]{hyperref}
\usepackage[ruled,linesnumbered]{algorithm2e}
\usepackage{setspace}
\usepackage{multirow}
\usepackage{multicol}
\usepackage{url}
\usepackage{tikz}
\usepackage{color}
\usepackage{adjustbox}
\usepackage{graphicx}
\usepackage{makecell}
\usepackage{multirow}
\usepackage{enumitem}
\usepackage{microtype}
\usepackage{algorithmic}
\usepackage{textcomp}
\usepackage{array}
\usepackage{dblfloatfix}
\usepackage{subcaption}
\usepackage{pgfplots}
\usepackage{tikz}
\usepackage{pifont}

\usepackage{amssymb}
\usepackage[figuresright]{rotating}
\usepackage{colortbl}
\usepackage{tabularray}
\usepackage{multirow}
\usepackage{longtable}
\usepackage{hhline}
\usepackage{caption}
\usepackage{subcaption}
\usepackage{rotating}
\usepackage{tabularray}
\usepackage[none]{hyphenat}

\begin{document}

\begin{frontmatter}





\author[add1]{Amir Fathalizadeh}
\author[add1]{Vahideh Moghtadaiee$^*$}
\ead{v$\_$moghtadaiee@sbu.ac.ir}
\author[add2]{Mina Alishahi}
\address[add1]{Cyberspace Research Institute, Shahid Beheshti University, Tehran, Iran}
\address[add2]{Department of Computer Science, Open Universiteit, Netherlands}


\title{A Survey and Future Outlook on Indoor Location Fingerprinting Privacy Preservation }




\begin{abstract}
The pervasive integration of Indoor Positioning Systems (IPS) arises from the limitations of Global Navigation Satellite Systems (GNSS) in indoor environments, leading to the widespread adoption of Location-Based Services (LBS) in places such as shopping malls, airports, hospitals, museums, corporate campuses, and smart buildings. Specifically, indoor location fingerprinting (ILF) systems employ diverse signal fingerprints from user devices, enabling precise location identification by Location Service Providers (LSP). Despite its broad applications across various domains, ILF introduces a notable privacy risk, as both LSP and potential adversaries inherently have access to this sensitive information, compromising users' privacy. Consequently, concerns regarding privacy vulnerabilities in this context necessitate a focused exploration of privacy-preserving mechanisms. In response to these concerns, this survey presents a comprehensive review of Indoor Location Fingerprinting Privacy-Preserving Mechanisms (ILFPPM) based on cryptographic, anonymization, differential privacy (DP), and federated learning (FL) techniques. We also propose a distinctive and novel grouping of privacy vulnerabilities, adversary models, privacy attacks, and evaluation metrics specific to ILF systems. Given the identified limitations and research gaps in this survey, we highlight numerous prospective opportunities for future investigation, aiming to motivate researchers interested in advancing ILF systems. This survey constitutes a valuable reference for researchers and provides a clear overview for those beyond this specific research domain. 
{To further help the researchers, we have created an online resource repository, which can be found at \href{https://github.com/amir-ftlz/ilfppm}{https://github.com/amir-ftlz/ilfppm}.}
\end{abstract}


\end{frontmatter}

\section{Introduction}\label{sec:intro}
Localization is the process of determining the spatial coordinates or position of a tracked item with respect to several reference points within a predefined area~\cite{location-definition}. There is an increasing demand for offering Location-Based Services (LBS) to individuals worldwide. 
The appearance of Global Navigation Satellite Systems (GNSS) was a turning point in localization, but lately, the traditional scope of positioning systems evolved with the advent of mobile computing, including the Internet of Things (IoT) and wearable devices, moving from military tracking and civilian navigation to location information. Location data has become essential in connecting the real and digital worlds for both personal and commercial uses. Nowadays, due to the widespread use of mobile devices, users are constantly generating various location data during their daily activities. Accordingly, there is a growing interest in LBS, which offers services depending on the user's location. Such services assist people in finding the appropriate route to different destinations, selecting various points of interest (POI) based on their preferences, and even remembering to take precautions for their healthcare purposes~\cite{s21031002}. 

All these LBSs require reliable and real-time location information, which is often accomplished using GNSS technologies. However, due to the severe attenuation of satellite signals and the low visibility of satellites in indoor environments, they mostly fail for indoor localization purposes~\cite{ASAAD2022109041}. The main indoor localization technologies are alternative signals such as Wi-Fi, Bluetooth, FM, AM, GSM, or LTE~\cite{6819800, RahmanMD17}. During localization, they should face indoor area challenges such as non-line-of-sight (NLOS) problems, multipath propagation, signal blockage, intentional and unintentional interferences, etc.

\begin{figure} [t]
    \centering   
    \includegraphics[width=0.7\textwidth]{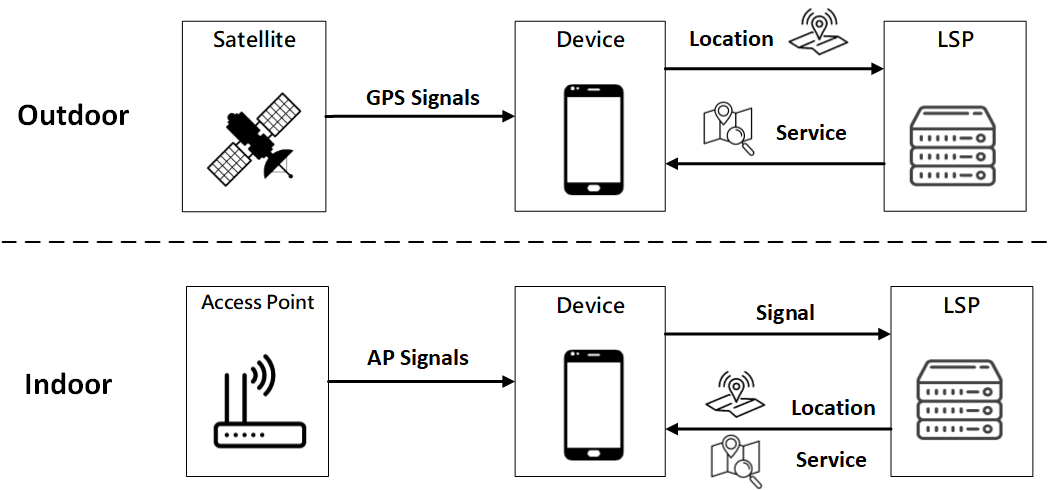}
    \caption{Outdoor vs. indoor positioning systems (no privacy is considered). 
    }
    \label{fig:indoor-outdoor}
    \vspace{-0.5cm}
\end{figure}

Indoor positioning has gotten so much attention in the last decade as individuals spend 80\% of their time indoors~\cite{KLEPEIS2001}, which shows the significant role of indoor localization. Unlike in outdoor positioning systems, the user has no direct access to her location, and it is calculated by the Location Service Provider (LSP), the server in the Indoor Positioning System (IPS), based on the signal information received from the user, as shown in Fig.~\ref{fig:indoor-outdoor}. Additionally, indoor localization demands higher precision compared to outdoor localization, which typically only requires meter-level accuracy. The indoor environments are diverse, including residences, workplaces, shopping malls, warehouses, hospitals, and retail centers.
Additionally, there are various applications for end users with different requirements for accuracy and coverage. For instance, while law-enforcement applications with urban/rural coverage have an accuracy of a few meters, ambient assisted-living applications require room-level coverage with an accuracy below $1m$~\cite{mautz}. Therefore, no IPS has evolved as an all-encompassing solution due to the unique characteristics of indoor settings and the diversity of applications. The coverage and accuracy are important to consider when selecting the base positioning technology, and the deployment and maintenance costs are crucial. The diverse solutions are a clear indicator that there is no single alternative for GNSS in indoor areas and different positioning technologies co-exist.

Therefore, one of the main obstacles to the widespread adoption of IPS in the world is privacy. 
When a user is linked to the network, IPS always "\textit{knows}" where they are. Without the user's prior consent, tracking her location is prohibited as users are generally reluctant to share their whereabouts. Location data reveal private and sensitive information on each user, including their health conditions, their interests, their views and behaviors, their usage amount of electricity, their workplaces, and their home and job locations. The leakage of private assembly line information, sensitive process information, and other relevant information are additional security risks posed by IPS in commercial or high-profile buildings. Moreover, the radio map of the buildings can be used to access the structural configuration, which might raise terrorist attacks and other negative actions. Therefore, practical answers are required to these privacy problems~\cite{MLSurvey-navit}. Authorities will establish stringent but pertinent norms and regulations for industries to use users' location data as nontechnical solutions. Without the user's explicit consent, data should not be shared with third-party organizations or people. Therefore, the lack of comprehensive efforts on this topic motivates us to examine indoor location privacy from various points of view thoroughly. 

\subsection{Existing Surveys and Our Contributions}
Several papers are surveying IPS in terms of accuracy, localization techniques, and technologies~\cite{MLSurvey-navit,MLSurvey-nessa,survey-zafari,10.1145/3396302}. Considering the security and privacy of IPS, on the other hand, there are two limited surveys~\cite{survey-1,survey-2} and only one comprehensive survey~\cite{SARTAYEVA2023103293}. The comparison of these three papers with our paper is reported in Table~\ref{tab:survey-comparison}, using labels such as Y (Yes), N (No), and L (Limited) to indicate the extent to which each survey covers specific subject.


{The first review paper in 2020,~\cite{survey-1}, conducts a limited analysis of privacy in IPS using the Preferred Reporting Items for Systematic Review and Meta-Analysis (PRISMA) guidelines~\cite{prisma}. The paper categorizes existing research papers (up to 2020) into three main groups: privacy on the device, during transmission, and at the server. The authors identify several privacy challenges specific to each category, such as the risk of unauthorized data collection on user devices, the interception of sensitive data during wireless transmission, and the vulnerabilities associated with centralized storage solutions, which could become targets for data breaches or misuse. 
However, as this research was conducted in 2020, it does not account for advancements in privacy-preserving techniques such as federated learning (FL), differential privacy (DP), or new cryptographic protocols that have since emerged.}

{The other survey in 2023~\cite{survey-2} starts by examining existing reviews on IPS and analyzes the technologies and techniques applied to them while discussing their advantages and disadvantages. 
This paper focuses on recent developments and surveys of a restricted number of studies— 15 papers published between 2018 and 2022— mostly related to privacy-preserving techniques in indoor location fingerprinting (ILF), most of which utilize secure two-party computation and homomorphic encryption. Although these cryptographic techniques are discussed in depth, the survey lacks a broader categorization of privacy threats and attacks, focusing primarily on encryption-based solutions without much emphasis on alternative methods such as anonymization, DP, or FL. Moreover, the paper does not take into account more recent research papers introducing novel privacy-preserving methods. Additionally, it does not discuss the potential vulnerabilities of these cryptographic techniques, leaving out a discussion on their performance trade-offs, scalability, or potential exposure to specific privacy attacks. Despite covering some technical challenges for IPS, this survey misses the opportunity to address privacy in a broader, evolving landscape of IPS technologies.}

Also in 2023, a comprehensive survey~\cite{SARTAYEVA2023103293} provides a broad overview of security and privacy issues in indoor positioning, encompassing a range of methods such as proximity-based, geometric, and collaborative techniques. 
It categorizes indoor positioning methods into two major categories: (1) non-collaborative methods, which include proximity-based methods, geometry methods, location fingerprinting, and others, and (2) collaborative methods, which include mobile proximity-based methods, mobile geometric methods, and others. For the sake of privacy and security, the survey applies the confidentiality, integrity, and authenticity (CIA) trilogy to both collaborative and non-collaborative methods. 
While their work presents a valuable analysis of security threats (e.g., jamming, replay attacks, data tampering), our paper takes a specialized approach by focusing exclusively on ILF privacy, an area of growing importance as ILF becomes more prevalent in real-world applications. One of the main differences in our work is the identification and detailed analysis of privacy-vulnerable entities specific to ILF systems. Unlike the broader scope in~\cite{SARTAYEVA2023103293}, which touches lightly on privacy concerns, our survey breaks down the unique risks each of ILF systems' entities faces, offering a more granular examination of how privacy can be compromised in fingerprinting-based systems. Additionally, our paper offers a thorough discussion on location privacy attacks specifically targeting ILF through signal fingerprints. These attacks, which exploit the inherent characteristics of ILF data, pose distinct threats that are not fully explored in these three existing surveys. In terms of solutions, while authors in~\cite{SARTAYEVA2023103293} focus on general security and privacy mechanisms, our paper thoroughly introduces advanced privacy-preserving mechanisms specifically for ILF. These include various cryptographic approaches, anonymization, DP, and FL, each critically evaluated for its effectiveness in ILF contexts. We also address the practical challenge of balancing accuracy and privacy in these systems— an essential consideration for real-world deployments, where maintaining system utility without sacrificing privacy is crucial. 

{\fontsize{7.2}{8.7}\selectfont
\begin{table*}[t]
\caption{Comparison of existing surveys on indoor localization privacy (\textbf{Y}: Yes, \textit{N}: No, L: Limited)}
\label{tab:survey-comparison}
\centering
\small
\begin{tabular}{lcccc}
\toprule
\textbf{Subject} & \textbf{\cite{survey-1}} & \textbf{\cite{survey-2}} & \textbf{\cite{SARTAYEVA2023103293}} & \textbf{Our survey} \\ 
\midrule 
ILF Focus           & \textit{N} & \textbf{Y} & \textit{N} & \textbf{Y} \\ 
Applications of IPS & \textit{N} & \textbf{Y} & \textbf{Y} & \textbf{Y} \\ 
ILF Entities' Privacy Vulnerabilities & \textit{N} & \textit{N} & \textit{N} & \textbf{Y} \\ 
Adversary Models  & \textit{N} & \textit{N} & \textbf{Y} & \textbf{Y} \\ 
Categorization of Privacy Attacks & \textit{N} & \textit{N} & \textit{N} & \textbf{Y} \\ 
Privacy Metrics & \textit{N} & \textit{N} & \textit{N} & \textbf{Y} \\ 
Datasets        & \textit{N} & \textit{N} & \textit{N} & \textbf{Y} \\ 
Cryptographic Techniques for ILF  & L & \textbf{Y} & L & \textbf{Y}\\ 
Anonymization Techniques for ILF  & \textit{N} & \textit{N} & L & \textbf{Y} \\ 
Differential Privacy (DP) for ILF & \textit{N} & L & L & \textbf{Y} \\ 
Federated Learning (FL) for ILF   & \textit{N} & \textit{N} & \textit{N} & \textbf{Y} \\ 
Comparative Analysis of Privacy Techniques & \textit{N} & \textit{N} & \textit{N} & \textbf{Y} \\ 
Case Studies on Privacy Violations in ILF & \textit{N} & \textit{N} & \textit{N} & \textbf{Y}\\ 
Challenges and Future Directions & L & \textit{N} & \textbf{Y} & \textbf{Y} \\ 
\bottomrule
\end{tabular}
\end{table*} }

Our paper, as mentioned, focuses on privacy-preserving mechanisms in the context of ILF, the most widely used localization technique in IPS. Incorporating the latest research and newly developed methods for addressing privacy concerns, this survey provides a comprehensive review of existing privacy-preserving solutions for ILF systems, including an in-depth analysis of privacy involving the entities in the IPS, the metrics and datasets commonly used, and the various attacks these systems face. Note that throughout the rest of the paper, when referring to IPS, we specifically mean IPS utilizing location fingerprinting as its localization technique.
The main contributions of this survey paper are outlined as follows:

\begin{itemize}
    \item To the best of our knowledge, this is the first deep dive survey considering indoor location fingerprinting privacy-preserving mechanisms (ILFPPM) from various novel aspects: indoor location privacy definitions, applications, entities vulnerabilities, possible attacks, privacy metrics, privacy protection techniques, challenges, and vision of future research directions.

    \item In addition to discussing IPS fundamentals, our exploration includes a thorough analysis of the various applications of IPS, investigating their practical implementations, and addressing the privacy concerns associated with the collection, storage, and utilization of location data in indoor spaces.

    \item We comprehensively identify privacy vulnerabilities and all sources of privacy leakages in ILF systems for the first time. We also examine them from diverse perspectives encompassing system entities, indoor location data structure, and information inference. Additionally, we introduce a novel categorization of adversary models and privacy attack models specific to these systems, marking the first systematic classification regarding privacy in ILF systems.

    \item To facilitate empirical investigations into ILFPPM, we provide a list of datasets from prior research. We also categorize all metrics (regarding localization, quality of service (QoS), and privacy) employed in existing studies for the first time, which is intended to establish a robust benchmark within the research community, fostering future empirical analyses and enhancing technical insights into ILF systems.

    \item Through the literature analysis, we have outlined the unresolved challenges and suggested various potential paths for future research in ILF privacy attacks and ILFPPM. 
\end{itemize}

\begin{figure*} 
\centering    \includegraphics[width=0.99\textwidth]{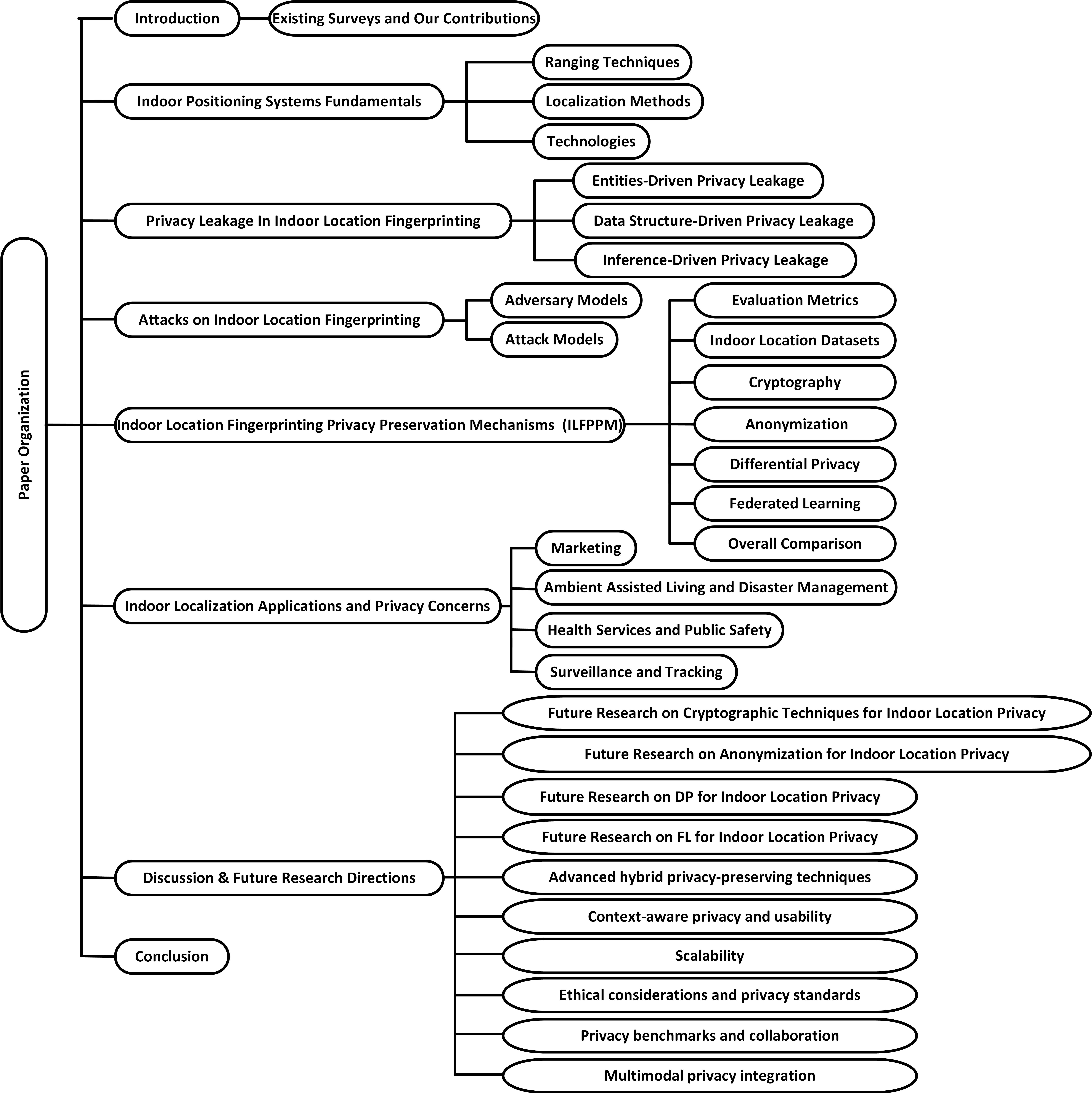}
    \caption{Taxonomy of the paper. 
    }
    \label{fig:ips_outline}
\end{figure*}

As outlined in Fig.~\ref{fig:ips_outline}, the rest of the paper is organized as follows. Section~\ref{sec:ips} presents IPS fundamentals including methodology, techniques, and technologies. Section~\ref{sec:locpriv} discusses the sources of privacy leakages in ILF, followed by the adversary and attack models in Section~\ref{sec:attack}. The existing ILFPPM are fully reported in Section~\ref{sec:ilppm}. All the datasets and metrics utilized in these studies are also presented in this Section. Section~\ref{sec:app} explains applications of IPS with their privacy concerns. Section~\ref{sec:Dis} discusses the future research directions, and Section~\ref{sec:conclusion} concludes the paper.
The abbreviations used in this paper are also listed in~Table~\ref{tab:abbr}.

\section{Indoor Positioning System Fundamentals}\label{sec:ips}
Understanding the fundamentals of IPS is essential before delving into privacy protection methods. IPS is a GPS-free system that continuously predicts the positions of people or objects in indoor environments by initially applying a distance estimation algorithm, followed by the localization algorithm~\cite{4343996}. Using a suitable ranging technique, an IPS calculates the distance between the target and anchor nodes whose coordinates are known a priori during the distance measurement phase. Using these distance observations, the IPS then uses several localization/positioning techniques to approximate the target's location.

\begin{table}[t]
	\centering
 	\caption{Summary of important abbreviations. }

	\resizebox{0.99\textwidth}{!}{
		\begin{tabular}[t]{ll ll}
			\toprule
			\textbf{Notation} &  \textbf{Description} & 	        
                \textbf{Notation}  &  \textbf{Description}  \\
			\midrule
			IPS & Indoor Positioning System & 
                ILF & Indoor Location Fingerprinting \\
                LPPM & Location Privacy Preserving Mechanism &
                ILFPPM & Indoor Location Fingerprinting Privacy Preserving Mechanism \\  
			LBS & Location Based Service &
                LSP & Location Service Provider \\
                RP & Reference Point &
                AP & Access Point \\
                RSS & Received Signal Strength &
                CSI & Channel State Information \\
                AoA & Angle of Arrival &
                PoA & Phase of Arrival \\
                ToA & Time of Arrival &
                TDoA & Time Difference of Arrival \\
                PoI & Point of Interest &
                IoT & Internet of Things \\
                ML & Machine Learning &
                NLOS & Non-Line Of Sight \\
                DP & Differential Privacy &
                CDP & Centralized Differential Privacy \\
                LDP & Local Differential Privacy &
                FL & Federated Learning \\
                MPC & Multi-Party Computation &
                2PC & Two-Party Computation \\
			\bottomrule
	\end{tabular}	}
	\label{tab:abbr}
\end{table}%

\subsection{Ranging Techniques}
In this section, we explore the various ranging techniques that have been used for indoor localization up to this point.

\subsubsection{Received Signal Strength (RSS)}
In general, Received Signal Strength (RSS) represents the intensity of a signal measured at the receiver, typically at a certain distance from a signal source. It is one of the most straightforward signal metrics to measure, providing valuable insights into the strength of wireless signals. However, particularly in indoor environments, RSS measurements can be susceptible to various factors that introduce inaccuracies. These factors include fading, shadowing, refraction, scattering, and reflections within complex indoor layouts.

The long-distance path-loss model has been frequently utilized for generating RSS values in an indoor area, as it demonstrates how RSS measurements vary with distance from an AP~\cite{Terzis}. The RSS values sensed from the $j$'th AP, ${S}_{u,j}$(dBm) at a physical distance $d_{j}$ in meter, are calculated as follows:

\begin{equation}\label{Path}
	{RSS}_{u,j}={s_{d_0}}_j - 10\alpha_{j}\log\left(\frac{d_j}{d_0}\right) + X_{\sigma_j},
\end{equation}

where ${s_{d_0}}_j$dBm is the power received at $d_0$m (usually $1$m) from $j$th AP, and $\alpha_{j}$ is the path-loss exponent associated with that AP. Due to the shadowing in indoor environments, $X_{\sigma_j} \sim \mathcal{N}(0,\sigma_j)$ models path-loss variation at a single point and is assumed to be a zero mean Gaussian random variable with a standard deviation given by $\sigma_j$dB.

Due to the sensitivity of RSS measurements to these environmental influences, they often yield erroneous distance estimates in IPS. To mitigate the fluctuations and inaccuracies associated with RSS, various techniques and machine learning (ML) approaches have been adopted~\cite{6843305}. These methods aim to enhance the reliability and precision of RSS-based positioning. Some of the strategies employed include the implementation of sophisticated filters and the utilization of ML algorithms. These approaches are designed to filter out noise, account for signal variations, and extract meaningful information from RSS data, ultimately improving the performance and robustness of IPS.

\subsubsection{Signal Propagation Time}
The distance between the target and the anchor node is also calculated based on the signal propagation time. These methods are typically more accurate than the RSS method. Time of Arrival (ToA) and Time Difference of Arrival (TDoA) are well-known distance calculation methods.
ToA-based algorithms require synchronizing between nodes, and TDoA methods were proposed to tackle it. Since TDoA takes into account the synchronization of the transmitters, it can somewhat address the problem of synchronization inaccuracy \cite{tdoa}. However, the performance of the ToA/TDoA-based systems is considerably hampered by the NLOS propagation of the signal \cite{4212819}. 


\subsubsection{Phase and Angle of Arrival}
The Angle of Arrival (AoA) and Phase of Arrival (PoA) approach estimate position by utilizing the angle and the phase that a signal makes with an antenna array \cite{8533724}. This improved ranging method is used to measure both the angle and the distance. The need for antenna arrays, which makes this approach complex and expensive, is one of its limitations \cite{Sakpere2017ASS}. The signal's timing of arrival at certain antenna elements may also be used in this approach, although this requires much more complicated hardware and precise calibration.

\subsubsection{Channel State Information}
Channel State Information (CSI) is another advanced method for calculating distances in indoor positioning. Unlike RSS, which provides only the amplitude value of a received signal, CSI offers a more precise estimation of the received signal's characteristics across the entire signal bandwidth. This improvement in accuracy is achieved by capturing the channel's frequency response received by each antenna, making CSI a valuable tool for distance estimation. While CSI often requires multiple antennas to gather this information effectively, it can be employed in a variety of localization systems, including both range-based and range-free approaches~\cite{csi-rss}. One notable advantage of CSI is its ability to provide not only signal amplitude but also phase information, enhancing the richness of data available for positioning calculations."

\subsection{Localization Methods}
The localization methods which are commonly used for indoor localization are listed below:

\subsubsection{Multilateration and Trilateration}
With the aid of three or more known nodes and the corresponding associated distances, it is a method for determining the position of the unknown node \cite{8457226}. Only three well-known nodes are employed in trilateration, which is a specific instance of multilateration. The target node's location in a two-dimensional space is determined by the intersection of three fictitious circles. However, because of the NLOS effect, which results in significant positioning errors, these circles do not converge at a single location in the real-world indoor environment. 

\subsubsection{Triangulation}
When the angle of arrival is known, it can be applied for positioning precision. It requires at least two anchor nodes and is moderately precise while being less complex \cite{POMARICOFRANQUIZ2014236}. The accuracy of the AoA estimation is crucial to the technique's location accuracy. The performance of localization can be improved by increasing the number of anchor nodes.

\subsubsection{Fingerprinting}
\begin{figure} [t]
    \centering
\includegraphics[width=0.7\textwidth]{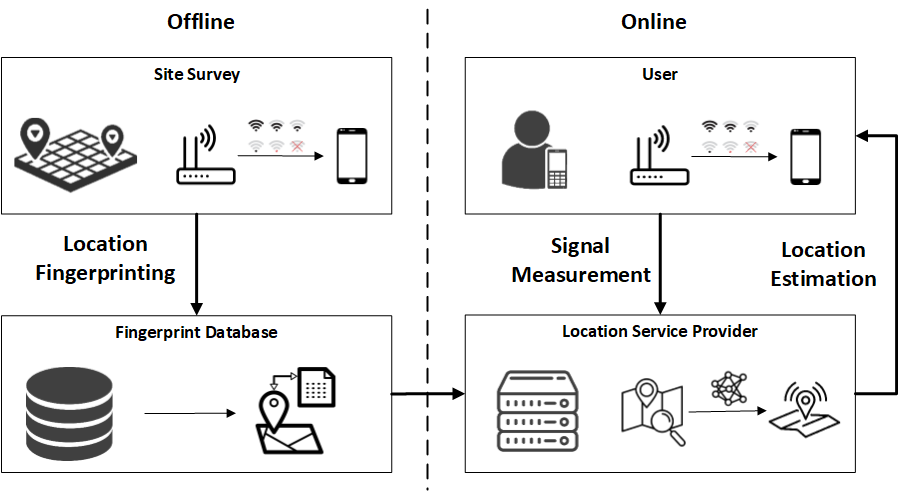}
    \caption{The flow of ILF technique.}
    \label{fig:fingerprint}
    \vspace{-0.5cm}
\end{figure}

Fingerprinting is a widely used indoor location technique that utilizes Wi-Fi, BLE, FM, AM, and ZigBee wireless access technologies \cite{8110721,zigbee,6819800,RahmanMD17,7103024}. As shown in Fig.~\ref{fig:fingerprint}, two phases are involved in the fingerprint-based localization method: offline and online. A radio map is created using the measured data for each recorded position during the offline phase using RSS or CSI data acquired at access points (APs) for various known indoor points called reference points (RPs).
The Received Signal Strength (RSS) values from all existing APs at known Reference Points (RPs), along with the corresponding $(x,y)$ coordinates of RPs, are stored in the radiomap within LSP. The recorded radiomap is characterized by a dimension of $M \times N$, where $M$ and $N$ represent the number of RPs and APs, respectively. The fingerprint at a given location $(x_i,y_i)$ is denoted as ${rss}_{i}^{RP} = [{rss}_{i1}, {rss}_{i2}, \ldots ,{rss}_{iN}]$, where ${rss}_{ij}$ represents the RSS of ${AP}_j$ at the \textit{i}th RP.
Moving to the online phase, the user's location fingerprint is obtained as a vector of RSS values, $S_u = [{RSS}_{u1}, {RSS}_{u2}, \ldots, {RSS}_{uN}]$, where ${RSS}_{uj}$ signifies the RSS of $AP_j$ received by the user. This vector is then transmitted to LSP for localization purposes. During this phase, the user's RSS vector is compared with the previously stored RSS values in the radiomap, and using machine learning methods, LSP estimates the user's position, $(x,y)$. 
If more accurately obtained offline data is used to build the radio map, this method offers great accuracy. However, it takes a great deal of work to create the radio map for large-area deployment (e.g., manpower, time, and cost). Additionally, for dynamic networks, the offline database needs to be regenerated whenever a node's position, even a single node, is unexpectedly changed or removed. 

ML algorithms are frequently used to improve fingerprinting accuracy and aid in building radio maps \cite{MLSurvey-nessa}. The simplest algorithm used for fingerprint-based localization techniques is k-Nearest Neighbors (kNN). This algorithm computes a distance metric that measures the separations between the target's measurements at various APs and the measurements taken during the training phase. Euclidean distance is the most widely used measurement of distance. The radio map's k closest RPs to the target with the smallest distances are chosen in this algorithm. The target's position is then inferred by averaging these RPs' coordinates. However, the kNN algorithm can also use alternative distance measures such as the Manhattan, Mahalanobis, and Minkowski distances~\cite{6817920}.

\subsubsection{Range-free}
Instead of calculating the position based on distance, range-free localization methods were proposed:
\begin{itemize}
    \item Centroid: Instead of using distance or angle measurement, this method estimates the location of the unknown node using a geometric relation. Once a reliable communication link has been established between each anchor node and the unknown target node, the positions of the anchor nodes are known. The centroid of the geometric shape formed by the position of the anchor nodes connected to the target node is taken to be the location of the unknown nodes. In \cite{range-free}, a Weighted Centroid Localization (WCL) approach-based BLE beacon IPS was suggested.
    \item Distance Vector Hop (DV Hop):
    In a multi-hop setting, this method involves estimating the distance vector depending on the hop count. An information table is kept up to date with the coordinates of the \textit{i}th node and the minimum hop count value from the anchor node to the \textit{i}th node. The anchor node broadcasts the position data to the nearby nodes, which subsequently broadcast it to more nodes, and so on. Finding the hop size for a specific hop is a crucial issue for this strategy. The distance between the node that is m hops distant from the anchor node is easily estimated after obtaining the average hop size \cite{range-free}.
\end{itemize}

\subsection{Technologies}
In this subsection, radio frequency wireless technologies that are most commonly used in indoor localization are briefly presented.
A diverse array of wireless technologies is employed to enable precise tracking and positioning of objects, devices, or individuals within indoor environments. The choice of technology depends on factors such as accuracy requirements, infrastructure availability, power efficiency, and the specific application domain. In this subsection, two prominent categories of such technologies which are radio frequency (RF) and non-radio frequency wireless technologies are briefly discussed.

\subsubsection{Radio Frequency}
Radio frequency (RF) technologies are frequently deployed for indoor localization due to their versatility and robust performance in various settings. Among the widely utilized RF technologies are Wi-Fi-based localization systems, which capitalize on the ubiquity of Wi-Fi access points in indoor spaces. These systems employ signal strength and triangulation techniques to estimate the position of Wi-Fi-enabled devices such as smartphones or tablets~\cite{wifi-indoor}. Bluetooth Low Energy (BLE) beacons also play a crucial role, especially in asset tracking and proximity-based applications. BLE beacons emit low-power signals that can be detected by smartphones or dedicated receivers, facilitating accurate indoor positioning~\cite{ble-indoor}. Ultra-wideband (UWB) technology, recognized for its high precision, is gaining traction in applications demanding centimeter-level accuracy, such as industrial robotics and healthcare~\cite{uwb-indoor}. FM/AM infrastructures, from another perspective, can be used to deploy various indoor localization methods without the need for additional hardware, which is applicable for long-range solutions with susceptibility to interference ~\cite{6819800, RahmanMD17}.

\subsubsection{Non-Radio Frequency}
Conversely, non-radio frequency wireless technologies are emerging as alternatives to RF solutions. Infrared (IR) technology relies on the transmission of light signals in the infrared spectrum to determine the location of objects equipped with IR transmitters and receivers. Although less common than RF technologies, IR systems offer advantages in specific scenarios, such as indoor positioning within controlled environments or line-of-sight requirements. Additionally, visible light communication (VLC) leverages LED lighting infrastructure to transmit data and positional information through modulated light signals. VLC systems have demonstrated potential for indoor localization while providing energy-efficient illumination~\cite{non-rf-indoor}.

\section{Privacy Leakage In Indoor Location Fingerprinting}\label{sec:locpriv}
This section provides an in-depth examination of the ILF system emphasizing privacy vulnerabilities and leakages. We offer a comprehensive understanding of the system entities and the underlying data structure, particularly in employing location fingerprinting techniques for indoor localization. Through this, we derive insights into potential areas of privacy leakage within these systems. Additionally, we extend this section to a detailed exploration of the broader privacy leakage arising from the information inferred about the users in ILF.

\subsection{Entities-Driven Privacy Leakage }\label{sec:entity}

\begin{figure}
    \centering
\includegraphics[width=0.9\textwidth]{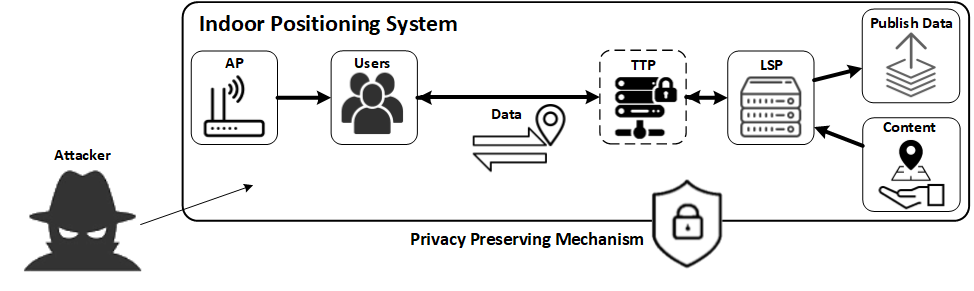}
    \caption{Privacy-vulnerable entities in ILF systems. }
    \label{fig:outline}
    \vspace{-0.3cm}
\end{figure}

Here, we introduce and analyze the fundamental components associated with the processes of data transmission and service provisioning in the ILF system so that we can comprehend the potential privacy vulnerabilities. As shown in Fig.~\ref{fig:outline}, the entities include Access Points (APs), Users, Trusted Third Parties (TTPs), Location Service Providers (LSPs), Content Providers, and Data Publishers. For each entity, we first introduce their characteristics, followed by the inherent privacy leakages within each entity, and then the potential privacy vulnerabilities that these entities may introduce. 

\subsubsection{Access Point (AP)}
The first hop in a data transmission often involves communication networks, mostly wireless APs. Wi-Fi, BLE, UWB, and RFID are among the applicable APs for indoor localization and data transmission. APs facilitate the transmission and reception of wireless signals within their coverage area. Devices seeking location information, such as smartphones or IoT devices, communicate with these APs through wireless signals. APs provide the means for wireless devices to connect to a network, enabling communication between the devices and the central network or server. This connectivity is essential for the exchange of location-related data during the localization process.

The potential privacy leakage for APs is the disclosure of their locations. The transmission of RSS/CSI data by APs to users during the service stage can potentially expose APs' locations through various methods. The disclosure of this data for APs, particularly through RSS, makes them susceptible to malicious activities, leading to physical access and potential damage. The distance from an AP can be calculated using RSS information, and multilateration and triangulation techniques can be applied to pinpoint the AP's location. Researchers have also found that the ratio of signal powers between the first peak and second peak in CSI can be used as a physical layer metric to estimate the distance from the AP~\cite{Crypt-Wang2015}. 

APs can be also curious and pose a privacy threat themselves. They are indispensable in IPS, but they have less power than a malicious localization server. The user collects one measuring value from each AP during localization. Theoretically, localization requires at least three measuring values for a 2-dimensional area, (or four for a 3D area) which are located in diﬀerent places. If the communication during measurement is unilateral, the measuring result is inherently secure against curious APs. Take RSS as an example, RSS values are measured at the user device by evaluating the signal strength from each AP without sending any information back. This kind of measurement leaks no information to the APs. If the communication during measurement is bilateral, colluding APs can obtain the full measuring result of the user and compute the location with the help of the localization server. Hence, making the measurements unlinkable or anonymous for measuring techniques involving bilateral communications is crucial. 
{From another perspective, when transmitting data, the user's device will automatically connect to the nearest AP for communication. As a result, this AP becomes aware of the user's proximity, which can inadvertently reveal the user's location to the AP, potentially compromising privacy. This connection process implies that even without detailed data exchange, the mere act of connecting provides the AP with information about the user's presence in its vicinity.}

\subsubsection{Users}
Users, the primary recipients of LBSs, are the central focus. The majority of LBSs are accessible via mobile devices. However, with the advancement of electronic technologies, wearable devices are also used for location applications and serving users. The data transmitted or received by users mostly contains sensitive information about their locations, jobs, habits, and POIs, emphasizing the need for robust privacy safeguards.

The privacy of the users is threatened in three different ways during the localization process. Firstly, the user may unintentionally reveal her location when she requests localization services, exposing her RSS/CSI information to attackers. Secondly, an untrusted party can capture the RSS/CSI measurements sent by the user, and use it to retrieve the user’s location from the server. Also, considering the server as an untrusted party, the server can determine the user's location based on her RSS/CSI measurements. 

The user itself can also act as a malicious party. Malicious users' actions in indoor localization are twofold: 1) Malicious users can access unauthorized information on the server through security misconfiguration and hacking or capture other users' data to track other users or sell the information for proﬁt. The localization server and localization process should protect this data from being exposed; 2) Collision of malicious users is another threat. Peers subscribing to the same LBS can either collude to launch attacks, or one adversary can create fake peers to obtain the information they seek. 
{For example, malicious users might create three fake anchor locations and use their corresponding distances to the target in an iterative trilateration process, based on a localization algorithm, to infer a target's location. This technique can mislead the localization system into producing inaccurate results or revealing sensitive information by exploiting the false data provided by these fake anchors.}

\subsubsection{Trusted Third Party (TTP)}
TTP is an impartial and trustworthy entity that facilitates secure interactions and transactions between parties. In the context of privacy, a TTP is often involved in managing sensitive information, ensuring secure communication channels, and validating the identities of the parties involved.
In indoor localization, it acts as a server dedicated to location privacy responsibilities, encompassing procedures such as anonymization and encryption, and is often managed by the LBS server or an arbitrarily trusted third party. Investigating the ownership and operational aspects of this server is crucial for understanding the trustworthiness of privacy-preserving measures.

As a trusted entity, it can access the transmitted data of users, which lacks adequate privacy protection. Consequently, an untrusted party may intercept and access the transmitted data by exploiting vulnerabilities within the TTP. TTPs themselves may also misuse the information they handle for unauthorized purposes, leading to privacy violations. This could involve the unauthorized sharing or selling of sensitive data.

\subsubsection{LSP}
The Location Service Provider (LSP) or LBS server is integral to responding to user inquiries and processing information for service provision. It contains important information about the building including the building map and location fingerprint database. It also receives and processes user information to provide location-based services to users. Therefore, protecting all these data is essential.


The inherent privacy risk of the LSB server lies where an untrusted party can gather real RSS/CSI measurements from various locations or generate an extensive set of artificial RSS/CSI measurements resembling those obtained through fuzzing. Subsequently, this entity can seek localization assistance from the server using the acquired measurements. By meticulously documenting pairs of RSS/CSI and their corresponding locations, the untrusted party can construct a fingerprint database resembling that maintained by the server. This deceptive replication of the server's fingerprint database enables the untrusted entity to potentially manipulate location-based services, exploiting the similarity between the fabricated and authentic data to gain unauthorized access or compromise the integrity of the localization system.


From another point of view, the most formidable adversary within this context is the malicious localization server. In the typical functioning of a system, the ultimate determination of a device's location heavily depends on the localization server. This pivotal role grants the server not only access to the location information of all mobile devices within its scope but also the capability to construct an approximate trajectory for a specific device through the analysis of multiple localization queries over time. The compromise of this server, whether through a sophisticated hack or other illicit means, opens the gateway to extracting significant information. The consequences extend beyond only location data, encompassing potentially sensitive details, user profiles, and any other information processed or stored within the server's databases.

\subsubsection{Content Provider}
Some contents, including maps and POIs, are essential for LBS functionality. While some LBS providers employ a third party to deliver this service, others use their own content. For indoor localization, the content may include the building map and semantic information of each section or room. Privacy of this content is also important as it has information about different parts of the buildings and the service provider would not like it to be publicly available, mostly for sensitive data. Consider a hospital where the service provider is reluctant for users to see the sensitive parts of the building, such as the electrical facilities, clean rooms, etc. 

Content Providers, however, do not pose a privacy risk as their communication with the LSP is unilateral. However, there is a chance that an untrusted party adds unusable or inappropriate content to the system, which leads to poor servicing, and the users will stop using the system.

\subsubsection{Data Publisher}
After gathering user data, the collected dataset can be leveraged for various purposes. This data may include the building's heat map for crowded locations or the frequently used trajectories. Investigating the utilization and potential re-identification risks of this published data is crucial for safeguarding user privacy.

The attacker can link the data of the published dataset with other public datasets to identify or re-identify users or other unauthorized activities.
Similar to the Content Provider, Data Publishing itself cannot introduce a privacy threat to the IPS, as the Data Publisher's communication with the server is unilateral.


\subsection{Data Structure-Driven Privacy Leakage}
In this section, inherent privacy vulnerabilities in the data structure of the ILF systems are introduced.
The location data in LBSs varies in indoor environments due to the different methods used for indoor localization. Different from outdoor positioning systems, where the main data is the coordinates (e.g. longitude and latitude), the data for indoor environments employing the fingerprinting technique is various signal characteristics vectors. As shown in Fig.~\ref{fig:loc-data}, the data in ILF is transmitted through the network as the tuple $<ID, Time, location > $~\cite{wernke-locpriv}. The $ID$ and $Time$ data can be defined the same for indoor and outdoor localization, but the $location$ data has a different type. The terms are explained below:

\begin{figure} [t]
    \centering    
    \includegraphics[width=0.7\textwidth]{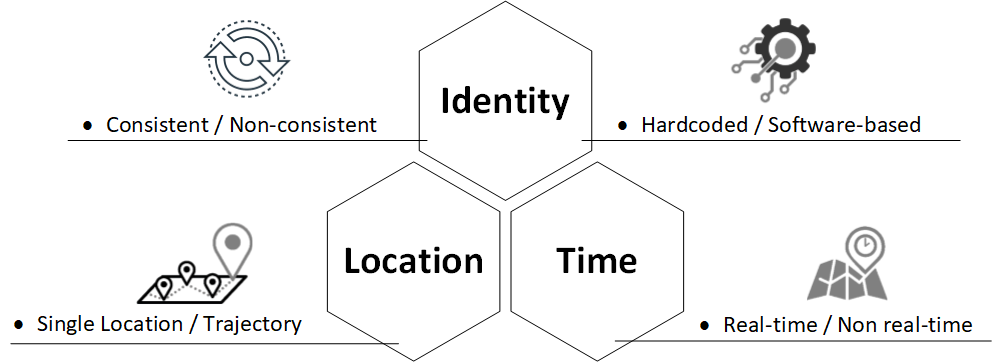}
    \caption{Data structure in ILF systems. }
    \label{fig:loc-data}
\end{figure}

\subsubsection{ID}

Identity refers to a user’s name, email address, or any other characteristic that distinguishes one individual from another. In the context of LBSs, identities can be categorized as either consistent or inconsistent~\cite{spatial-tra}. Consistent identities are those that remain the same across different sessions or interactions with the service, such as a fixed email address, phone number, or a hardcoded identifier like a MAC address or IMEI. These identifiers allow LBSs to consistently recognize and track the same user over time. Inconsistent identities, on the other hand, can vary between sessions or interactions. For instance, users might employ LBSs under a pseudonym or a temporary, self-defined ID provided by the application, such as a session ID or a custom identifier created for a specific purpose. This variability in identity helps protect user privacy by making it more challenging to link activities or track individuals across different sessions.

Identity-driven privacy leakage, commonly called deanonymization, aims to identify a user by gathering information. These leaks include:
\begin{itemize}
    \item
    \textit{Single identity leaks} or \textit{personal identification assaults} involve pinpointing a user's identity through their indoor location, such as within a faculty building's lab, or indirectly by narrowing down identification possibilities—such as inferring gender, educational level, or access to restricted areas. This approach allows for further investigation to potentially uncover someone's identity~\cite{id-attack1}. For instance, a solitary presence in a room, like a teacher's office, offers a higher chance of exact identification, while a crowded space, like a student-filled lab, lowers the probability of specific identification but doesn't eliminate it.
    \item \textit{Multiple identity information leaks}, \textit{meeting disclosure leaks}, or \textit{aggregated presence assaults} try to determine the relationship between two or more people, or between people and an aggregated property, such as whether people meet at a particular time or the approximate number of persons visiting a lab~\cite{8329504}. As a case in point, when two or more people go to the same lab at specific time intervals, they are highly probably co-workers or friends, or they have a meeting at that time.
\end{itemize}

\subsubsection{Location} The primary tool for determining the location of an object is spatial information. Coordinates, such as longitude and latitude, are employed for outdoor locations, while signal characteristics are used for indoor localization. Other information, like a store's name, may also be associated with a location. In indoor fingerprinting localization, hence, the location data involve the characteristics of the existing signal in the indoor area, such as RSS or CSI when RF technologies are utilized for localization, as mentioned in Section~\ref{sec:ips}. This location data is received by the user's device, sent to the server for the calculation of the user's location, and the estimated location is then sent back to the user.
Location-driven privacy leakage refers to locating crucial areas, such as a person's exact workplace in an indoor environment~\cite{7568598}. 
    
 \subsubsection{Time} 

Some LBSs integrate identification and location information, along with a timestamp for each location. Data transmission can be categorized as either real-time or non-real-time. Certain LBSs opt for non-real-time data collection, saving information for later use, while others prioritize real-time services, each presenting unique privacy challenges. Real-time privacy protection is more complex due to the need for scalable solutions that can handle dynamic and unpredictable user movements. Achieving global optimization in this context refers to the challenge of effectively managing and analyzing real-time data across various locations and users to ensure accurate and efficient service delivery while maintaining privacy. This is particularly difficult because it requires balancing the need for precise, up-to-date information with robust privacy safeguards. Two examples of LBSs that utilize real-time location data are navigation systems and augmented reality games. 

Based on the combination of time and location data, spatial information can be broadly categorized into two groups: Single and Trajectory. Single location information refers to static spatial data about a specific point, such as the layout of a room or the coordinates of a stationary object. In contrast, trajectory information captures the movement and path of an object over time, used for tracking people or assets for navigation, asset management, or security purposes.

Time-driven privacy leakage itself may only lead to limited information, but when it is combined with location data it reveals sensitive information. This combination can be categorized as follows:

\begin{itemize}
    \item Location along with discrete time determines whether a user is present at a location at a specific time, where the presence or absence of the user may disclose information.
    \item Location along with continuous time, or simply tracking, is to establish a user trace by putting all or part of the sequence of events together. Stalking is also the term used to describe this type of leak.
\end{itemize} 

In summary, identity, location, and time trilogy constitute the sensitive data in IPS; thus, they are the main targets of the adversaries. This data is also considered dynamic, as it can change rapidly over time.
Regardless of the data format and the dynamic nature of it, utilizing an LBS generates a substantial amount of location-related data. This data is highly correlated, as datasets of real-world places often exhibit significant coupling relations, and frequent connections between positions may provide more information than initially anticipated. 
However, an attempt to gain access to each identity, location, and time separately might not be helpful for the attackers. For example, accessing the identity by itself may not reveal anything meaningful, but when combined with the location and/or time, it can result in the disclosure of valuable information.

\subsection{Inference-Driven Privacy Leakage}
In ILF systems, there are also some privacy leakages arising from the information inferred about the users~\cite{inference_2024}.
Two critical aspects can be considered as the main dimensions that can be inferred and violate privacy in indoor fingerprinting localization: identity inference and profiling completeness. Identity inference raises concerns about unauthorized tracking and monitoring, unveiling sensitive details through IPS. Simultaneously, profiling completeness, while enhancing user profiles, poses risks by holding sensitive information. These two privacy dimensions are explained below in the following subsections~\cite{8329504}. 

\subsubsection{Identity inference} 
Identity inference from indoor location data introduces multifaceted privacy concerns. Location tracking through IPS not only unveils sensitive details about a user's activities, habits, and interests but also poses the risk of unauthorized tracking and monitoring without user consent~\cite{id-attack1}. The identification capabilities of these systems, based on unique location patterns, raise ethical concerns related to targeted marketing, personal profiling, and potential discriminatory practices. Furthermore, the prospect of re-identification, where seemingly anonymous users can be linked to their identity by combining location data with external information, underscores the need for stringent privacy safeguards. Additionally, location data can lead to potential inference that extends beyond exact location disclosure, as frequent visits to specific places can reveal significant information, such as health conditions. Collectively, these aspects highlight the intricate challenges associated with preserving identity privacy in the context of location data and emphasize the importance of ethical considerations and robust protective measures. 

\subsubsection{Profiling completeness} 
Profiling completeness involves the frequent identification of Points of Interest (POIs) within a building based on user locations~\cite{poi-indoor}. Utilizing this information becomes crucial for gaining a deeper understanding of user behavior in real-world scenarios and predicting future actions. This data, commonly employed for marketing and surveillance purposes, possesses the capability to significantly enhance the comprehensiveness of profiles compared to basic data alone. The functionality of location data to seamlessly "connect the dots" allows for the automated generation of profiles for individuals or organizations. IPS, pivotal for tracking a user's position within a building, holds sensitive information. Unauthorized access to this information could lead to tracking and monitoring without user awareness or consent, echoing the challenges discussed in identity inference. 


\section{Attacks on Indoor Location Fingerprinting} \label{sec:attack}
An adversary attacks IPS aiming to collect the location information of users and use it for their benefit. This section discusses the attacks on ILF systems, investigating the different aspects including adversary and attack models to obtain meaningful data about the user's identity, location, and time. 

\subsection{Adversary Models}
In general, user privacy means that no passive adversary (including a curious or malicious server) can determine the location of an honest user after intercepting all protocol messages. Server privacy, however, protects the server's data from being compromised by malicious users using location queries~\cite{Crypt-Yang2018-2}.
To discuss threat models on privacy preservation in indoor localization, we need to explain privacy assumptions on how entities can behave in IPS. Four assumptions are taken into consideration which are explained in the following subsection. 

\subsubsection{Fully trusted} In this setting, all parties involved are trusted to follow the protocol and not deviate from the agreed-upon behavior. This setting assumes that all participants will faithfully execute the protocol and will not attempt to gain unauthorized information or compromise the data being processed. This setting provides the lowest level of assurance in terms of privacy and requires a high degree of trust among the involved parties, which is very rare. 

\subsubsection{Semi-trusted (honest-but-curious)} In this setting, it is assumed that the parties involved will adhere to the protocol, but they may attempt to gain additional information from the protocol execution. In other words, while they will not intentionally deviate from the protocol, they may analyze the exchanged signals to infer additional information about the inputs or computations of other parties. Protocols designed for the Semi-trusted setting aim to prevent participants from learning more than they are supposed to, even if they try to gain information through analysis. 

\subsubsection{Fully malicious} In this setting, it is assumed that one or more parties may actively deviate from the protocol in an attempt to subvert its goals or compromise the privacy of the data. This setting represents the most adversarial scenario, where participants may collude, attempt to cheat, or launch attacks to learn more than they are allowed to or to undermine the privacy guarantees of the protocol. Protocols designed for a fully malicious setting typically incorporate stronger security mechanisms to withstand active attacks and attempts at manipulation. 

\subsubsection{Uunilateral-malicious} Introduced in~\cite{Crypt-Yang2018-2}, the unilateral-malicious setting is weaker than the fully malicious setting but stronger than the conventional Semi-trusted setting. We formulate the malicious behaviors specific to user sessions in the unilateral-malicious setting, such as manipulating Wi-Fi fingerprints and disclosing locations. We expect the server to act fairly honestly. In other words, while the server may be interested in learning the location of a user, it should run the protocol instance honestly to offer a high-quality service. 
{
Since a server providing inaccurate or unreliable location services would likely be noticed by users and lead them to stop using the service, the risk of deliberate poor service is lower. In other words, if the server consistently fails to show correct locations, users would abandon it, reducing the incentive for such attacks. However, the server faces more difficulty in detecting malicious behavior by users.
}

Understanding the difference between various settings lies in the level of trust and cooperation expected from the participants, as well as the defenses and security measures needed to ensure privacy and integrity in the face of potential adversarial behaviors and threats.

\begin{figure} [t]
    \centering
    \includegraphics[width=0.85\textwidth]{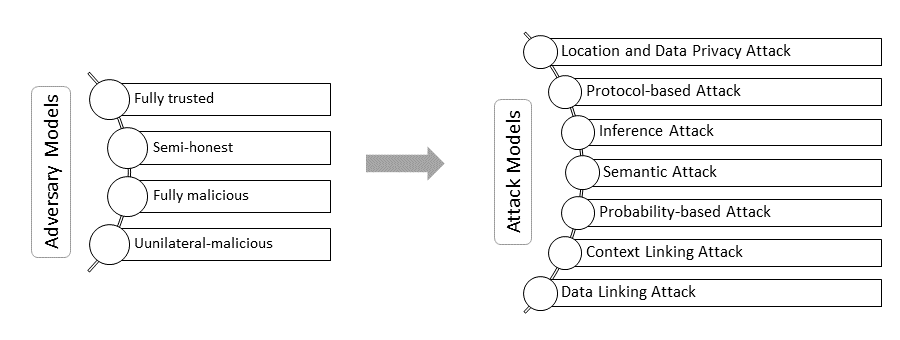}
    \caption{Adversary and attack models overview in ILF systems.}
    \label{fig:attack}
    \vspace{-0.3cm}
\end{figure}

\subsection{Attack Models}
In the context of computer security and privacy, an attack model refers to a representation of potential threats, adversaries, and their capabilities that a system or application may face. It helps security analysts, developers, and researchers understand and anticipate the possible ways in which a system might be compromised or exploited. Following the privacy assumptions, the attackers intend to seek data from IPS. As shown in Fig.~\ref{fig:attack}, this section aims to explain the attack models for IPS, and how attackers exploit entities' vulnerabilities and location fingerprinting data structure to obtain information about entities in IPS. 


\subsubsection{Location Privacy and Data Privacy Attacks}\label{sec:loc_data_priv}
Location privacy revolves around the potential threats to an individual's privacy resulting from the collection, use, and disclosure of their location information. Targeted advertising, monitoring daily habits, surveillance, stalking and harassment, discriminatory uses, and unauthorized access are examples of location privacy concerns. 
Overall, location privacy is a subcategory of data privacy that is disparate in importance. By this means, from the perspective of the user, different locations have different privacy requirements. For instance, the majority of people worry significantly more about keeping their houses and workplaces private than they do about revealing where they visited in a shopping center. Despite differing levels of data sensitivity, the dangers associated with location privacy remain a substantial concern~\cite{8329504}. 

\begin{figure}[t]
    \centering
    \includegraphics[width=0.9\textwidth]{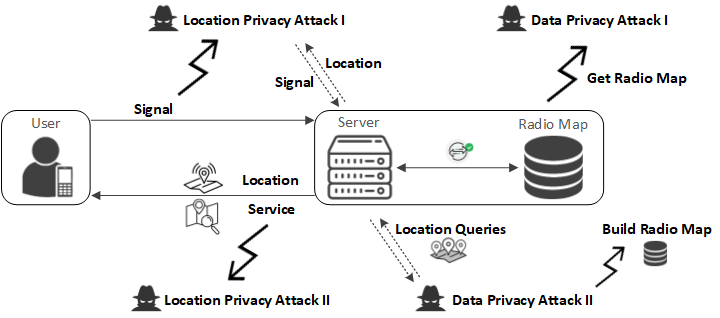}
    \caption{Location privacy and data privacy attacks in ILF systems.}
    \label{fig:threat_model}
\end{figure}

The major attack model on IPS is introduced in~\cite{Crypt-Li2014}, which is followed by most papers. As shown in Fig.~\ref{fig:threat_model}, this attack model categorizes the threats into the \textit{Location Privacy Attacks} and \textit{Data Privacy Attacks}. In both of them, location and data can be attacked in two ways as follows~\cite{Crypt-Li2014}: 

\begin{itemize}
    \item \textit{Location Privacy Attack I}: The attacker directly obtains the user’s location information from the query from the server.
    \item \textit{Location Privacy Attack II}: The attacker indirectly infers the user's location information by accessing the user’s WiFi RSS vector. 
    \item \textit{Data Privacy Attack I}: The attacker accesses the WiFi ﬁngerprint database stored in the server.
    \item \textit{Data Privacy Attack II}: The attacker builds a WiFi ﬁngerprint database, which is similar to the one stored in the LSP and the localization accuracy of using the first database is comparable with that of using the latter database.
\end{itemize}

The location privacy attack can be applied by passively capturing the signal on the communication channel between the user and the localization server, which can lead to important information. By either capturing user-to-server or server-to-user exchanged data, the attacker can get the location of the user. The data privacy attack can also occur by gaining direct access to data on the localization server via a hack or misconfiguration, or by querying patterned locations (e.g., querying for location on a patterned grid) to form the fingerprint database and consequently misuse it.

Note that another categorization with the same principle is applied by using terms \textit{user-side privacy attack} and \textit{server-side privacy attack}~\cite{Crypt-Yang2018}. 
{The localization server and the APs are regarded as adversaries in breaking the user’s location privacy. The user, on the other hand, is seen as an adversary to compromise the server’s database security.} 
The user must not reveal her location to untrusted parties, and the service provider should also keep its WiFi ﬁngerprint database from unauthorized leaking. For example, a malicious user may download the database and sell it for proﬁt.

\subsubsection{Protocol-based attack} In this attack, the malicious actor uses cryptographic methods to extract data from encrypted data or discovers misconfiguration in data transmission protocols between user and server. Specifically in the chosen fingerprint attack~\cite{Crypt-Yang2018} or known location attack~\cite{Crypt-Yang2018-2}, the user chooses special ﬁngerprints, such as all-zeros or single-one ﬁngerprints, to compromise the whole server’s database.


\subsubsection{Inference attack} 
An inference attack involves an adversary deducing sensitive information about individuals or systems by analyzing seemingly harmless or non-sensitive data. These attacks exploit patterns and relationships within the data to draw conclusions that may not be explicitly available. In the context of location-based services, inference attacks can reveal a person's activities, habits, or even identity by analyzing patterns in their location data over time.
Recent work in \cite{rftrack_inference} demonstrates how inference attacks in indoor location models can exploit wireless signal data, such as Received Signal Strength Indicator (RSSI) values, to infer a device's location without requiring physical access to the environment. By analyzing temporal signal strength patterns and employing techniques like Reinforcement Learning (RL), attackers can model device movements and deduce location information, posing significant privacy risks to indoor localization systems.
One specific type of inference attack is a membership inference attack (MIA), where an adversary attempts to determine whether a particular individual's data was used in training a machine learning model~\cite{shokri}. This form of attack has also been studied in indoor location models~\cite{mem_secrypt24}, where attackers analyze subtle cues from model outputs to infer whether specific location data was part of the training set, potentially compromising user privacy.

\subsubsection{Semantic attack}
In a semantic attack, the objective of the attacker is to obtain specific semantic details about an individual. Location data inherently contains semantic information, and the disclosure of such information can unveil personal details and behavioral tendencies of users. Within indoor location datasets, an adversary can deduce the purpose of a user's visits to particular locations by monitoring their movements or extended stays~\cite{FATHALIZADEH2022102665}. For example, in a hospital, the adversary may discern whether the user is visiting a specific doctor, and in a shopping center, they can identify specific shops where the user lingers. Additionally, this information might reveal whether the user is searching for or purchasing particular products in a shopping center, consulting a specific specialist in a hospital, or meeting a particular professor at a university.

\subsubsection{Probability-based attack} This attack type focuses on how the attacker would change his probabilistic belief about the sensitive information of a user after acquiring the captured data, rather than on what records, attributes, and tables the attacker may link to a target user. In general, this subset of privacy attacks seeks to achieve the uninformative principle, which aims to keep the gap between pre- and posterior beliefs as minimal as possible.

\subsubsection{Context linking attack} Most location-based attacks make use of some context. Correlating metadata from different sources, such as social media profiles, public databases, or online activities, allows attackers to link contexts and uncover additional details about individuals. When launching a localization attack, it is simple to combine contextual knowledge with the observed location data to determine a target's exact location. For instance, a personal context linking attack \cite{context-link} may be used to remove all the unnecessary areas from a user's location after reducing an obfuscated area to a specified spot. Contextual information can be used in conjunction with precise location data to carry out identity attacks. An attacker can assume that their target is or was in a special room of a hospital at a particular moment, for instance, if they know someone's affiliation and discover it on a hospital room list.

\subsubsection{Data linking attack} Attacks on the published indoor location datasets are considered ~\cite{FATHALIZADEH2022102665}: identification/re-identification and Record/Attribute/Table linkage attacks. The attacker's goal in the identification/re-identification attack is to identify or re-identify the victim in the whole dataset among all the other users that might be presented there. In an indoor location data set, an attacker with some background knowledge about an individual can identify whether he/she has been in a specific location. This is particularly important when spatial-temporal locations and trajectories are monitored by an adversary. A linkage attack occurs when an adversary can link a record owner \emph{(i)}  to a record in a published dataset, \emph{(ii)} to a sensitive attribute in a revealed dataset, or \emph{(iii)} to the published table of data by itself. These attacks are called \emph{record linkage}, \emph{attribute linkage}, and \emph{table linkage} attacks, respectively. Linkage attacks are characterized by the adversary's prior knowledge. In the case of record linkage and attribute linkage, the adversary knows that a specific individual's data is presented in a dataset and he wants to learn about sensitive information. Table linkage attack, on the other hand, focuses on understanding whether a known individual's information is available in the released dataset or not. The adversary could deduce sensitive information from the disclosed dataset based on the distribution of sensitive values in the group to which the individual belongs in an attribute linkage attack

\section{Indoor Location Fingerprinting Privacy Preservation Mechanisms (ILFPPM)}\label{sec:ilppm}
This section reviews and compares existing studies on ILFPPM. We first discuss the evaluation metrics and the datasets considered in these studies. Four major privacy preservation techniques have been utilized for  ILFPPM, including cryptographic methods, anonymization techniques, DP, and FL. Each of these methods is discussed in the following subsections, accompanied by a brief explanation of the privacy method, necessary definitions, the applicability of privacy methods to indoor settings, as well as their respective advantages and disadvantages.

Note that there are also passive solutions for localization~\cite{passive,passive-beacon}, in which the fingerprinting dataset is available for the users and the localization process is done locally on the user's device. In this method, however, the privacy of the user is preserved, but the server-side privacy, or in other words, server data privacy is not preserved, so it is not considered in this categorization.









\subsection{Evaluation Metrics}\label{sec:metric}
Here, we discuss various metrics used in the different studies to evaluate the ILFPPM. All these metrics and their associated research are mentioned in Tables \ref{tab:crypt}, \ref{tab:Anon}, \ref{tab:dp}, \ref{tab:fl}. We categorize all metrics into the three main groups: localization accuracy, Quality of Service (QoS), and privacy metrics. 

\subsubsection{Localization accuracy metrics}

The widely used metric in ILFPPM determines the localization accuracy. Location error metrics are crucial for evaluating the accuracy and precision of localization systems in various applications. These metrics are referred to by the following names: Location Error, Distance Error, Root Mean Square Error (RMSE), Mean Squared Error (MSE), Mean Absolute Error (MAE), Accuracy, Error Rate, Success Rate, Spatial Loss, RSS Loss, Cloaked Region (CR) Area, Area under the Success Rate curve (Area of ASR), Floor Detection, Kendall's Tau Distance, Control Value, Count, and Area.

\medskip
\textbf{1) Location Error} and \textbf{Distance Error} are basic metrics that measure the Euclidean distance between the estimated location and the actual location. These metrics help assess how far off the localization predictions are in real-world scenarios.

\medskip
\textbf{2) Root Mean Square Error (RMSE)} is the square root of the average of the squared differences between the predicted and actual locations. RMSE provides a comprehensive measure of the localization error by giving more weight to larger errors, making it useful for applications where large deviations are especially problematic.

\medskip
\textbf{3) Mean Squared Error (MSE)} is the average of the squared differences between predicted and actual locations. It penalizes larger errors more than smaller ones, making it effective in scenarios where minimizing larger errors is a priority.

\medskip
\textbf{4) Mean Absolute Error (MAE)} is the average of the absolute differences between the predicted and actual locations. It provides a straightforward measure of the average magnitude of localization errors, without penalizing larger errors more heavily than smaller ones, as MSE does.

\medskip
\textbf{5) Accuracy}, \textbf{Error Rate}, and \textbf{Success Rate} are metrics that represent the percentage of correctly localized instances. \textbf{Accuracy} measures the proportion of correct predictions, \textbf{Error Rate} quantifies the percentage of incorrect localizations, and \textbf{Success Rate} refers to the ratio of successful localization attempts (those within a certain error threshold) to the total number of attempts, offering a simple way to evaluate system performance.

\medskip
\textbf{6) Spatial Loss} measures the discrepancy between the predicted and actual locations in a spatial context, considering the geometric relationships between points. This metric is particularly important when dealing with spatial data, where the relative positions of locations are critical.

\medskip
\textbf{7) RSS Loss} evaluates the difference in the Received Signal Strength (RSS) values between predicted and actual locations. This metric is often used in localization systems that rely on wireless signal strength measurements to infer location.

\medskip
\textbf{8) Cloaked Region (CR) Area} refers to the size of the area in which a user’s location is obfuscated to preserve privacy. A larger CR area implies greater uncertainty about the user's precise location, which can enhance privacy but potentially reduce localization accuracy.

\medskip
\textbf{9) Area of ASR (Area under the Success Rate curve)} represents the total area under the curve that plots the success rate over varying error thresholds. It provides an aggregate measure of how well the system performs over a range of localization error tolerances.

\medskip
\textbf{10) Floor Detection Accuracy} is specific to indoor localization systems that operate across multiple floors of a building. It measures the system's ability to correctly identify the floor on which the user is located.

\medskip
\textbf{11) Kendall's Tau Distance} is a ranking metric used to evaluate the correlation between the predicted and actual orderings of locations. It is relevant in systems that rank possible locations or trajectories.

\medskip
\textbf{12) Control Value} refers to a metric that adjusts or controls for certain variables during localization, ensuring that the system's performance is assessed under standardized conditions.

\medskip
\textbf{13) Count} and \textbf{Area} metrics are more context-specific, with \textbf{Count} typically referring to the number of successful localization attempts or the number of messages exchanged, and \textbf{Area} referring to spatial areas of interest, such as the size of the localization region or the area affected by localization errors.

\subsubsection{Quality of Service (QoS) metrics}

The second group is employed to assess the Quality of Service (QoS) of the ILFPPM. It includes various metrics, such as QoS Loss, Communication Overhead, Time Cost, Run Time, Execution Time, Response Time, Energy Consumption, Entropy, Number of Messages, Noise Effect, Bandwidth Cost, and Battery Life.

\medskip
\textbf{1) QoS Loss} measures the degradation in service quality concerning predetermined thresholds, indicating the system’s ability to meet desired performance levels. A higher QoS loss indicates that the system is unable to maintain the required performance, especially under varying loads or adverse conditions. 

\medskip
\textbf{2) Computation and Communication Overhead} quantify the additional computational or communication resources required to execute tasks. This is particularly relevant in cryptography-based approaches, where encryption introduces extra processing requirements and data transmission overheads. These overheads, while protecting data privacy and security, increase resource consumption, potentially reducing efficiency and responsiveness. 

\medskip
\textbf{3) Time Cost} metrics encompass various aspects, such as \textbf{Execution Time}, which refers to the duration it takes to complete a specific task, and \textbf{Run Time}, which measures the total duration of system operation. These collectively shape the system’s responsiveness and real-time performance, indicating how well the system can function under different conditions or workloads. 

\medskip
\textbf{4) Response Time} refers to the time it takes for the system to respond to a query or input. In real-time systems, low response time is crucial for maintaining user satisfaction and ensuring the system is reactive to environmental changes. 

\medskip
\textbf{5) Energy Consumption} measures the amount of energy required to execute a task or keep the system running. This is a critical metric, especially for battery-powered devices, as excessive energy consumption can reduce device longevity and make the system impractical for mobile or IoT devices in continuous operation. 

\medskip
\textbf{6) Entropy} is a measure of uncertainty or randomness in the system. In the context of ILFPPM, it can refer to the degree of uncertainty in localization data or how well the system maintains data privacy by introducing uncertainty into location predictions to prevent inference attacks. 

\medskip
\textbf{7) Number of Messages} exchanged during communication sessions provides insights into network traffic and resource utilization. A high number of messages can increase communication overhead, impacting both system scalability and efficiency, particularly in environments with limited bandwidth or high latency. 

\medskip
\textbf{8) Noise Effect} metrics evaluate the impact of signal distortion or interference on data transmission and processing accuracy. Noise can come from various sources, such as physical obstacles, environmental interference, or network congestion. Assessing the noise effect is essential for ensuring robustness and system reliability, especially in challenging, noisy environments. 

\medskip
\textbf{9) Bandwidth Cost} refers to the amount of bandwidth consumed during data transmission. Systems that require frequent communication or transfer large amounts of data can impose significant bandwidth costs, which can affect both the user experience and network efficiency, especially in low-bandwidth scenarios. 

\medskip
\textbf{10) Battery Life} evaluates the longevity of devices operating in the system. Systems that consume excessive battery power may limit the practicality of ILFPPM, particularly for mobile or IoT devices, where energy efficiency is a key requirement. 

\medskip
\textbf{11) Adaptation speed} measures how quickly a model adjusts to new environments by either measuring the number of training steps needed to reach a target accuracy or assessing accuracy after a fixed number of steps. It is calculated as the inverse of these values, offering a way to evaluate how efficiently a model adapts to changes.

\subsubsection{Privacy metrics}

The last group consists of privacy-related metrics. Privacy metrics are essential for evaluating the effectiveness of privacy-preserving mechanisms in data handling, storage, and transmission. These metrics are Location Privacy, Probabilistic Metrics, and Normalized Central Penalty (NCP) in ILFPPM papers. 

\medskip
\textbf{1) Location Privacy Metrics} evaluate the degree to which a system preserves the location privacy of individuals when handling spatial datasets. These metrics often consider aspects like spatial granularity, where the system determines the precision of location data shared, and proximity disclosure, which measures the risk of revealing sensitive locations based on proximity to known points of interest. Higher granularity or more frequent proximity disclosures can lead to increased privacy risks, making these metrics vital for understanding the trade-offs between utility and privacy in localization systems. Techniques such as $k$-anonymity, location obfuscation, or cloaking are often evaluated using these metrics to determine their effectiveness in safeguarding location privacy. 

\medskip
\textbf{2) Probabilistic Metrics} are particularly important for privacy-preserving mechanisms based on DP. These metrics assess the likelihood that an adversary can infer sensitive information from seemingly anonymized or noisy data. In the context of DP-based indoor localization systems, probabilistic metrics measure the probability of identifying an individual or location through repeated queries or by analyzing the noise added to the location data. By quantifying the risks associated with probabilistic inference attacks, these metrics help in determining whether a system adheres to the privacy guarantees promised by differential privacy. They also highlight the trade-off between data utility (how useful the data is) and privacy protection, ensuring the system effectively balances the two. 

\medskip
\textbf{3) Normalized Central Penalty (NCP)} is a quantitative measure used to assess privacy violations in datasets, often applied to anonymized or perturbed data. It indicates how much sensitive information is leaked in a dataset by evaluating the centrality or proximity of sensitive records. Higher NCP values indicate greater privacy risks, showing that more sensitive data is potentially exposed. In the context of ILFPPM, NCP is used to analyze the extent to which individual locations or other sensitive attributes can be reidentified after anonymization or perturbation. By providing a numerical measure of privacy leakage, NCP allows researchers and developers to evaluate the effectiveness of privacy-preserving mechanisms and compare them across different systems. This is especially useful when determining the trade-off between privacy protection and system performance, helping to fine-tune privacy mechanisms to achieve optimal results. 

Other important privacy metrics include \textbf{Adversary Success Rate}, which measures the likelihood that an attacker can successfully breach the system’s privacy defenses. A lower adversary success rate corresponds to a more secure system. \textbf{Anonymity Set Size} measures how many individuals share the same privacy profile, with larger sets providing better anonymity guarantees. \textbf{Exposure Risk} evaluates the likelihood that sensitive information is exposed over time through repeated queries or location data sharing. Finally, \textbf{Linkability} measures the ability of an adversary to link anonymized records back to the original individual or location, thus breaching privacy. High linkability scores indicate a vulnerability in the anonymization or obfuscation techniques used by the system.

\subsection{Indoor Location Datasets}
Given the diversity of proposed methods and the fact that many of them may not be highly applicable to real-world datasets, most papers, even those intended for application to real-world datasets, employ their proposed methods in a simulated testbed. The open-source datasets widely used for indoor localization, even in non-privacy-focused papers, are UJIndoorLoc~\cite{UJIIndoorLoc} and JUIndoorLoc~\cite{JUIndoorLoc}. The CSUCY~\cite{Anon-Konstantinidis2015}, KIOSUCI~\cite{Anon-Konstantinidis2015}, CRAWDAD~\cite{crawdad}, PosData~\cite{DP-Wang2018}, SPAWC~\cite{SPAWC_dataset}, Geolife, and Gowalla~\cite{inference_2024} are also public datasets that are used in ILFPPM papers. In some papers, the dataset's location is not explicitly mentioned; instead, they mention the physical properties of the real-world dataset prepared for the proposed method. Some other datasets are Cyberspace Research Institute (CRI)~\cite{CRI-dataset}, Sangmyung University~\cite{DP-WookKim2018}, Rutgers University~\cite{DP-Sadhu2017}, Tampere University~\cite{tampere}, Colombia University~\cite{Crypt-Quijano2019}, University of Helsinki~\cite{Crypt-Nieminen2021}, University of Minho~\cite{minho_dataset}, and Guangzhou Xinguang shopping mall~\cite{Xinguang} datasets.

\subsection{Cryptography in ILFPPM} \label{Sec:Cry}
Here, we briefly explain the cryptographic methods and then discuss the studies in which these methods are used in the concept of ILFPPM. Finally, we provide guidelines on utilizing cryptography in ILFPPM.

\subsubsection{Cryptographic techniques}

\medskip
\textbf{Homomorphic encryption (HE)}
{Homomorphic encryption is a cryptographic technique that enables computations to be executed on encrypted data directly, without the need for prior decryption. The outcomes of these computations remain encrypted, and upon decryption, yield results identical to those obtained had the operations been conducted on the original, unencrypted data. Two well-known homomorphic cryptosystems used to protect the privacy of indoor location data are DGK~\cite{7d5894b} and Paillier~\cite{48910-X_16}. The DGK protocol, developed by Damgård, Geisler, and Krøigaard in 2007, is an efficient solution for the millionaire’s problem and supports homomorphic operations on small plaintexts. Similar to Paillier, DGK allows computations on encrypted data, but the plaintext space in DGK is smaller and can be chosen dynamically. The computations in DGK are performed modulo \(N\), whereas in Paillier they are performed modulo \(N^2\). The Paillier cryptosystem is an additive homomorphic cryptosystem, meaning that given the encryption of two values, \(m_1\) and \(m_2\), one can compute the encryption of their sum, \(m_1 + m_2\). Additionally, Paillier supports the multiplication of a ciphertext by a plaintext number.
}

\begin{figure}[t]
    \centering
    \includegraphics[width=0.8\textwidth]{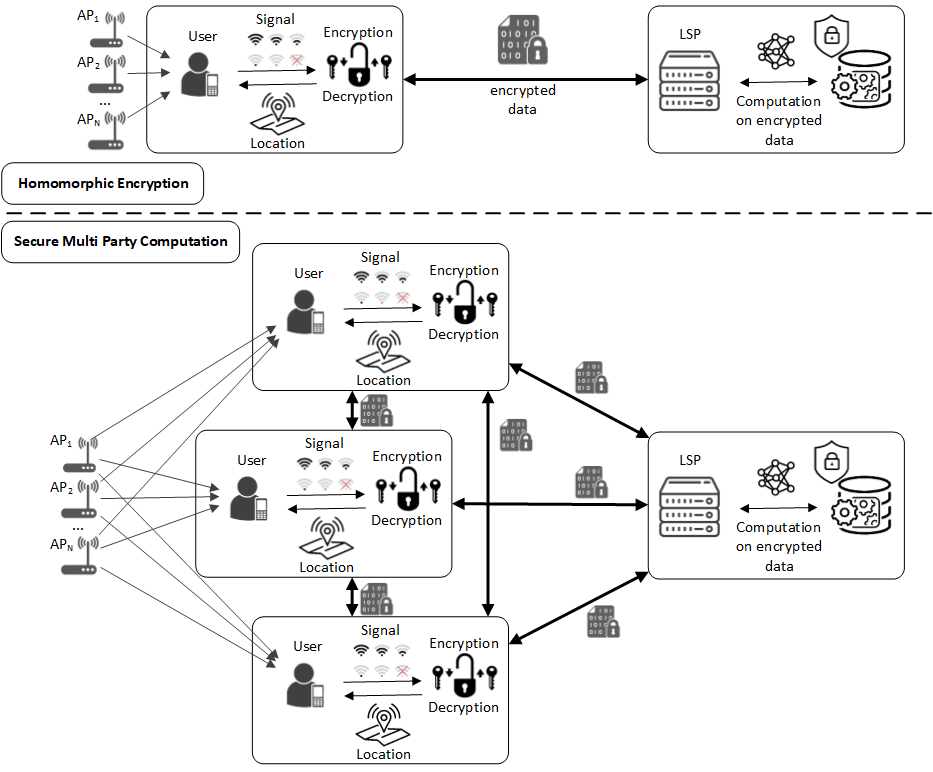}
    \caption{Homomorphic encryption vs SMPC in ILF systems. 
    }
    \label{fig:homo_mpc}
        \vspace{-0.3cm}
\end{figure}



\medskip
\textbf{Secure Multi-Party Computation (SMPC)}:
{Secure Multi-Party Computation (SMPC), or simply Multi-Party Computation (MPC), is a subfield within cryptography that aims to develop techniques enabling multiple parties to collaboratively compute a function using their inputs while maintaining the privacy of those inputs. MPC protocols can rely on either secret sharing or garbled circuits. In secret-sharing-based methods~\cite{359176}, the participating parties do not have specific roles, unlike the garbler-evaluator paradigm seen in Yao's protocol. Instead, the data associated with each wire is shared among the parties, and a protocol is employed to evaluate each gate. The function is defined as a \lq circuit' over a finite field, although it can also be defined over a ring or bits, as seen in the GMW protocol, which is implemented and further optimized in ABY~\cite{aby} and ABY2.0~\cite{aby2.0}. Such circuits referred to as \lq arithmetic circuits' in the literature, consist of addition and multiplication \lq gates', where the values being operated on are defined over a finite field.
}


\begin{table*}[t]
\caption{Studies on cryptography-based ILFPPM. 
}
\label{tab:crypt}
\centering
\scriptsize
\begin{tabular}{ccclclcl}
\toprule
\textbf{Paper}	& \textbf{Year}  & \textbf{Localization} & \textbf{Privacy Method} & \textbf{Dataset}  & \textbf{Metric} & \textbf{Adversary Model}  & \textbf{Attack Model}\\
\hline 

\cite{Crypt-Li2014}
& 2014
& kNN
& Paillier
& {\begin{tabular}{@{}c@{}} Simulated \\ \hline Private\end{tabular}} 
& {\begin{tabular}{@{}l@{}} Coms. Overhead \\ Loc. Error\end{tabular}}
& Fully Malicious
& Loc. \& Data Priv. Attack \\ \hline 

\cite{Crypt-Wang2015}
& 2015
& Fuzzy Logic
& Paillier
& CRAWDAD 
& Loc. Error
& Semi-trusted
& Loc. Priv. Attack \\ \hline 

\cite{Crypt-Zhang2016}
& 2016
& SVM
& Paillier
& Private 
& Coms. Overhead
& Fully malicious
& Loc. \& Data Priv. Attack \\ \hline 

\cite{Crypt-Yang2018}
& 2018
& -
& {\begin{tabular}{@{}l@{}} Paillier \\ Garbled Circuit \end{tabular}}
& - 
& -
& Semi-trusted
& {\begin{tabular}{@{}l@{}} Loc. \& Data Priv. Attack  \\ Protocol-based Attack \end{tabular}} \\ \hline 

\cite{Crypt-Yang2018-2}
& 2018
& -
& {\begin{tabular}{@{}l@{}} Paillier \\ Garbled Circuit \end{tabular}}
& Simulated 
& -
& Unilateral-malicious  
& {\begin{tabular}{@{}l@{}} Loc. \& Data Priv. Attack  \\ Protocol-based Attack \end{tabular}} \\ \hline 

\cite{Crypt-Richter2018}
& 2018
& kNN
& {\begin{tabular}{@{}l@{}} Paillier \\ Garbled Circuit \end{tabular}}
& {\begin{tabular}{@{}c@{}} Private \\ JUIndoorLoc \\ Tampere Uni \end{tabular}} 
& {\begin{tabular}{@{}l@{}} Loc. Error \\ Floor Detection\end{tabular}}
& - 
& Loc. Priv. Attack \\ \hline 

\cite{Crypt-Quijano2019}
& 2019
& kNN
& {\begin{tabular}{@{}l@{}} Paillier \\ DGK Algorithm \end{tabular}}
& Colombia Uni
& {\begin{tabular}{@{}l@{}} Execution Time \\ Battery Life\end{tabular}}
& - 
& - \\ \hline 

\cite{Crypt-Wang2019}
& 2019
& kNN
& {\begin{tabular}{@{}l@{}} ABY Framework \\ Arithmetic Sharing \\ Boolean Sharing \\ Yao's Sharing \end{tabular}}
& {\begin{tabular}{@{}c@{}} Simulated \\ \hline Private \\ same as \cite{Crypt-Li2014} \end{tabular}}
& {\begin{tabular}{@{}l@{}} Coms. Overhead \\ Run Time\end{tabular}}
& Fully malicious
& Protocol-based attack  \\ \hline

\cite{Crypt-Jarvinen2019}
& 2019
& kNN
& Trusted Third Party 
& Simulated
& {\begin{tabular}{@{}l@{}} Execution Time \\ Energy Consumption \end{tabular}}
& Fully trusted 
& {\begin{tabular}{@{}l@{}} Loc. \& Data Priv. Attack  \\ Protocol-based Attack \end{tabular}} \\ \hline 

\cite{Crypt-Eshun2019}
& 2019
& kNN
& {\begin{tabular}{@{}l@{}} Paillier \\ Spatial Bloom \end{tabular}}
& -
& Coms. Overhead
& Semi-trusted 
& Loc. Priv. Attack   \\ \hline

\cite{Crypt-Zhang2020}
& 2020
& kNN
& Paillier
& Private
& {\begin{tabular}{@{}l@{}} Coms. Overhead \\ Loc. Error \\ Time Cost \end{tabular}}
& Fully malicious 
& Loc. \& Data Priv. Attack   \\ \hline

\cite{Crypt-Wu2020}
& 2020
& kNN
& {\begin{tabular}{@{}l@{}}Expectation- \\ maximization \\ Bayes Network \end{tabular}}
& UJIIndoorLoc
& {\begin{tabular}{@{}l@{}} Coms. Overhead \\ Loc. Error \\ Time Cost\end{tabular}}
& Semi-trusted 
& Loc. Priv. Attack  \\ \hline

\cite{Crypt-Nieminen2021}
& 2021
& kNN
& {\begin{tabular}{@{}l@{}} Paillier \\ One Time Pad \\ Garbled Circuits  \end{tabular}}
& {\begin{tabular}{@{}c@{}} Simulated \\ \hline Uni of Helsinki \end{tabular}}
& {\begin{tabular}{@{}l@{}} Coms. Overhead \\ Loc. Error \\ Time Cost\end{tabular}}
& Semi-trusted 
& Loc. \& Data Priv. Attack  \\ \hline

\cite{Crypt-Beets2022}
& 2022
& kNN
& {\begin{tabular}{@{}l@{}} Arithmetic Sharing \\ Delta Sharing \\ Yao's Sharing \end{tabular}}
& Simulated
& {\begin{tabular}{@{}l@{}} Coms. Overhead \\ Run Time\end{tabular}}
& Semi-trusted 
& - \\ \hline

\cite{Crypt-Hu2022}
& 2022
& MLE
& Paillier
& Private
& {\begin{tabular}{@{}l@{}} Coms. Overhead \\ Loc. Error\end{tabular}}
& - 
& - \\ \hline

\cite{10414030}
& 2024
& kNN
& {\begin{tabular}{@{}l@{}} Inner Product \\ Encryption\end{tabular}}
& {\begin{tabular}{@{}c@{}} Simulated \\ \hline JUIndoorLoc \end{tabular}}
& {\begin{tabular}{@{}l@{}} Loc. Error \\ Coms. Overhead \\ Time Cost\end{tabular}}
&  Semi-trusted
& Data Priv. Attack \\ 

\bottomrule
\end{tabular}
\end{table*}

\subsubsection{Comparative analysis of Cryptographic techniques}
Cryptography-based ILFPPM uses encryption to protect users' indoor locations. Encryption is the most popular mechanism, probably because it is a standard solution for securely transmitting data. Fig.~\ref{fig:homo_mpc} illustrates how Homomorphic encryption and SMPC are utilized in ILF systems.

Table~\ref{tab:crypt} reports all the studies that employ Cryptography for ILFPPM. Authors in~\cite{Crypt-Li2014} presented the first indoor location privacy preservation work by encrypting measured RSS. It also introduces the attack models for data privacy and location privacy mentioned in section~\ref{sec:loc_data_priv}, which other studies have followed. However, the proposed method was found insecure in~\cite{Crypt-Yang2018}. Additionally, Järvinen et al. introduced PILOT~\cite{Crypt-Jarvinen2019}, the first efficient solution for privacy-preserving indoor localization using STPC, where the computational load is outsourced to two semi-trusted servers to achieve practical performance.

Most papers rely on the k-Nearest Neighbors (kNN) algorithm for the localization process, as highlighted in the Table; however, alternative methods have also been explored. For instance, Support Vector Machine (SVM)~\cite{Crypt-Zhang2016} has been employed for enhanced classification accuracy in localization tasks, while fuzzy logic~\cite{Crypt-Wang2015} has been applied to handle uncertainty and imprecision in probabilistic location estimation. Additionally, Maximum Likelihood Estimation (MLE)~\cite{Crypt-Hu2022} has been utilized to improve the accuracy of location predictions by maximizing the probability of the observed data. These approaches provide more flexibility in modeling complex environments, particularly in scenarios where the signal measurements exhibit significant variability or noise. For the data applied in simulations and experiments, most papers primarily use RSS values for localization; however, the integration of cryptographic techniques with CSI data has also been explored, as demonstrated in~\cite{Crypt-Wang2015}. Commonly used metrics to assess the effectiveness of these proposed schemes include the computation and communication overhead, as well as the localization error. These metrics provide a comprehensive evaluation of both the security and performance trade-offs, ensuring that the cryptographic enhancements do not excessively degrade system efficiency while maintaining accurate location estimates.

The use of cryptographic techniques in privacy-preserving indoor location data has been explored in various ways across the surveyed studies, with most leveraging the Paillier cryptosystem, Garbled Circuits (GC), and the DGK algorithm. Paillier encryption is favored for its homomorphic properties, enabling encrypted data computations without decryption, which is commonly employed in studies such as \cite{Crypt-Li2014}, \cite{Crypt-Wang2015}, and \cite{Crypt-Zhang2020} to maintain location privacy with minimal computational overhead. However, its performance trade-offs, particularly in real-time localization systems, include latency and communication overhead, as noted by \cite{Crypt-Wu2020} and \cite{Crypt-Nieminen2021}. Similarly, Zhang et al.~\cite{Crypt-Zhang2016} present a privacy-preserving indoor localization (PPIL) system based on SVM and Paillier encryption, although it faces limitations due to reliance on semi-trusted servers, as discussed in~\cite{Crypt-Yang2018}.
In contrast, Garbled Circuits offer more robust security features and are typically applied in scenarios involving fully malicious adversaries. Studies such as \cite{Crypt-Yang2018} and \cite{Crypt-Richter2018} combine Garbled Circuits with Paillier encryption to strengthen privacy guarantees, especially when both location and data privacy are at risk. However, the higher computational costs of GC limit its suitability for low-latency applications, as noted by \cite{Crypt-Zhang2016} and \cite{Crypt-Beets2022}. Nieminen and Järvinen~\cite{Crypt-Nieminen2021} further explored the combination of Paillier encryption and Garbled Circuits in a PPIL scheme for the user-server setting. 

While cryptographic methods such as Paillier and Garbled Circuits provide varying levels of privacy and efficiency, the choice of technique depends heavily on the adversary model, the dataset size, and the performance requirements of the application.
A key observation across all studies is the diversity of datasets used, ranging from simulated environments to public datasets like JUIndoorLoc, Colombia University, and Tampere University. Studies based on real-world datasets, such as \cite{Crypt-Richter2018}, \cite{Crypt-Quijano2019}, \cite{Crypt-Wu2020}, and \cite{10414030}, provide more practical insights into the performance of these cryptographic methods in realistic settings. However, there is a need for more extensive testing on public datasets to standardize performance comparisons. 

Overall, cryptographic methods operate without relying on a TTP, but they often involve relatively expensive operations. While these methods can be adapted to both client-server and outsourced settings such as the outsourcing-to-two-servers model discussed in~\cite{kamara2011secure}, they typically require computationally heavy pre-computations in the setup phase for each localization query. These computations, especially on user devices, need significant resources, which can be a concern for devices with limited battery life.

\subsubsection{Discussion on employing Cryptography-based methods}
To apply cryptographic techniques in ILF systems, the following guidelines will help achieve robust privacy protection while maintaining operational effectiveness:

\medskip
\textbf{Balancing privacy and localization accuracy:} 
Methods like homomorphic encryption allow encrypted data processing without decryption but introduce significant computation delays. In scenarios where both precision and speed are critical in localization, SMPC can offer a more balanced approach. SMPC enables joint computations on private data while maintaining strong privacy guarantees and typically improves system responsiveness compared to fully homomorphic encryption. However, SMPC can introduce communication overhead, particularly in larger systems. Thus, it is important to evaluate the trade-off between acceptable precision loss, privacy protection, and the system’s ability to handle the computational and communication demands.

\medskip
\textbf{Adaptability to adversarial models:} When dealing with fully malicious adversaries, more robust cryptographic methods like Paillier encryption combined with advanced privacy protocols are needed, as used in secure multi-party computations. For less severe adversarial models, lighter cryptographic techniques such as homomorphic encryption or secret sharing schemes can be utilized to reduce computational overhead while maintaining an acceptable level of privacy and security. These methods provide a balance between efficiency and protection, depending on the severity of the adversarial model.

\medskip
\textbf{Encryption overhead and scalability:} Cryptographic methods introduce varying levels of computational and communication overhead, which may impact scalability in large systems. For instance, homomorphic encryption methods are more resource-intensive but offer higher privacy. Simpler techniques like secret sharing or lightweight secure multi-party computations are computationally lighter but may offer less robust protection. It is important to choose encryption schemes that align with the privacy requirements and scalability needs of the desired deployment.



\subsection{Anonymization  in ILFPPM}
Here, we briefly explain the anonymization techniques and then discuss the studies on anonymization-based ILFPPM. Finally, we provide guidelines on employing anonymization in ILFPPM.

\begin{figure} [t]
\begin{subfigure}{0.45\textwidth} 
\centering
\includegraphics[width=.85\linewidth]{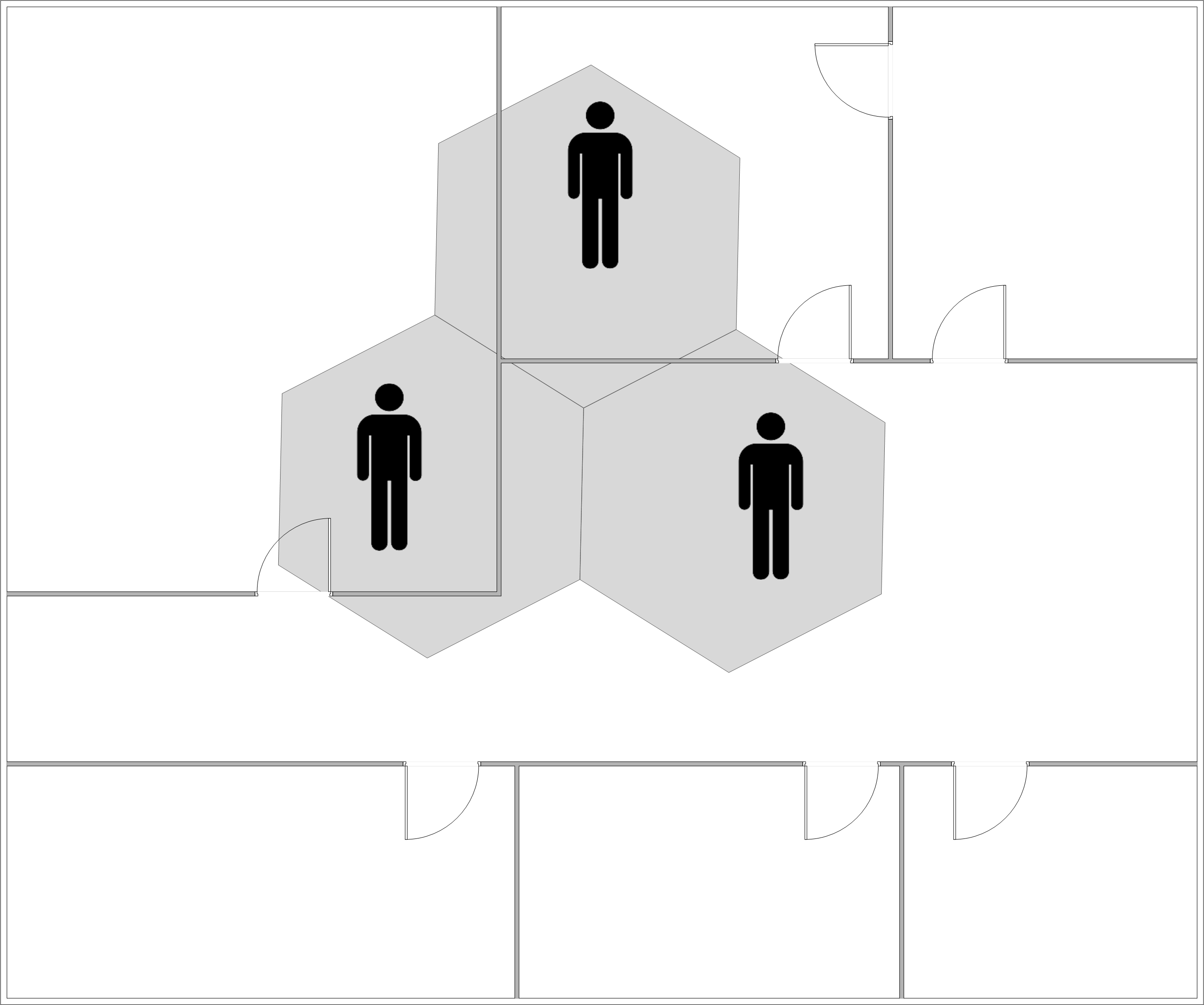} 
		\caption{location }
		\label{fig:k-anon-loc}
\end{subfigure}
\hfill
\begin{subfigure}{0.45\linewidth} 
\centering
\includegraphics[width=.85\linewidth]{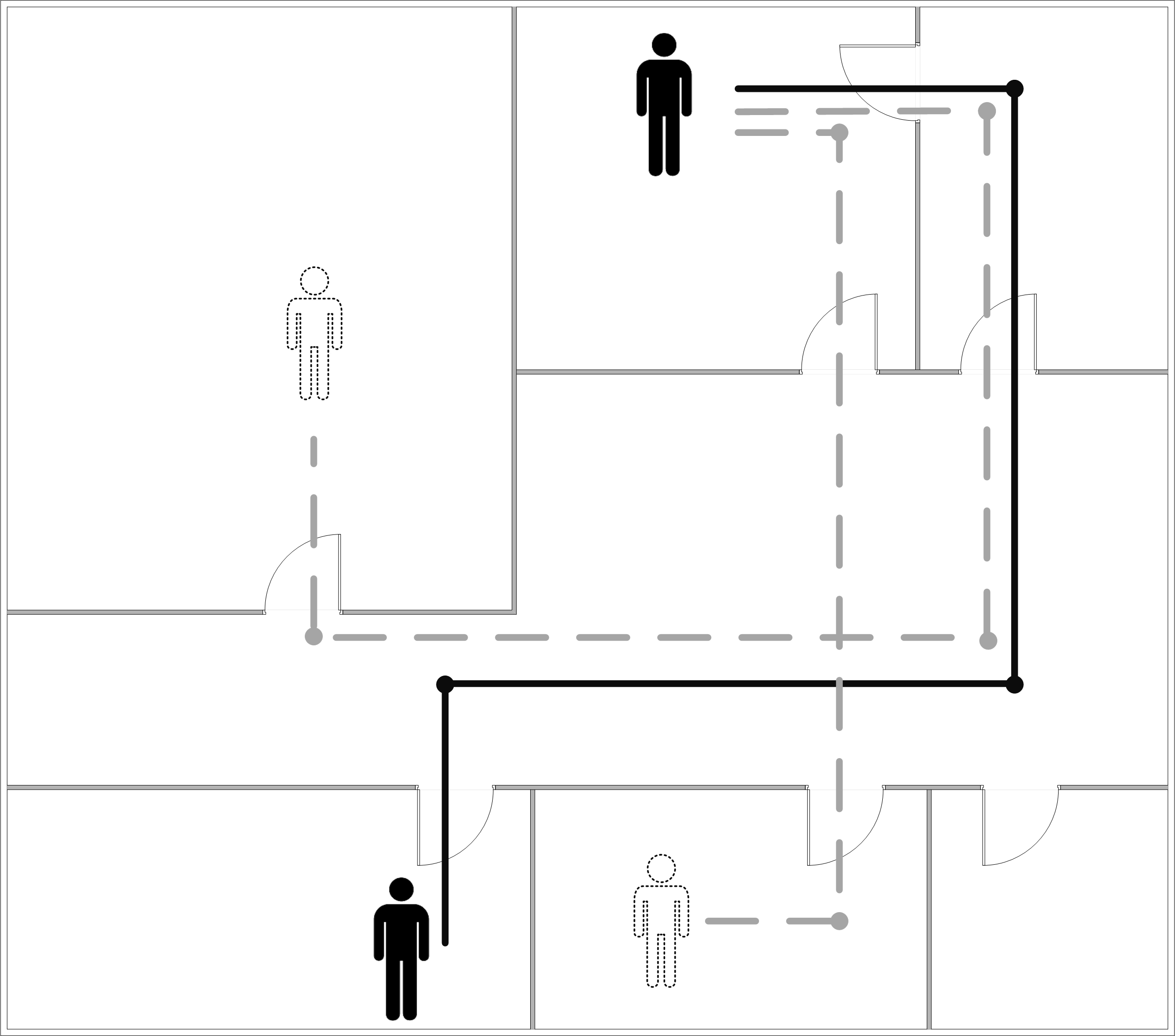} 
	\caption{trajectory}
	\label{fig:k-anon-tra}
\end{subfigure}
\vspace{-0.2cm}
\caption{Examples of $3$-anonymity for location and trajectory privacy protection}
\label{fig:k-anon}
\end{figure}

\subsubsection{Anonymization techniques}
Data anonymization is a fundamental technique in data privacy and security, aiming to protect individuals' sensitive information while still enabling the analysis and utilization of valuable datasets. Anonymizing data involves the transformation of personally identifiable information (PII) or sensitive attributes within a dataset in such a way that the identities of individuals are concealed, and their privacy is preserved. Several key anonymization techniques, including $k$-anonymity, $\ell$-diversity, and $t$-closeness, have been developed to achieve different levels of privacy protection and utility preservation in data sharing and analysis. These techniques are typically applied to quasi-identifiers, attributes that, when combined, can potentially lead to the identification of individuals.

\textbf{$k$-anonymity}~\cite{samarati1998protecting}:
    Given a dataset $\mathcal{D}$, a quasi-identifier $\mathcal{QI}$, and a parameter $k$, $\mathcal{D}$ satisfies $k$-anonymity if and only if, for each combination of quasi-identifier values in $\mathcal{QI}$, there exist at least $k$ records in $\mathcal{D}$ with identical quasi-identifier values.

\textbf{$\ell$-diversity}~\cite{1617392}:
    Given a dataset $\mathcal{D}$, a quasi-identifier $\mathcal{QI}$, and a parameter $\ell$, $\mathcal{D}$ satisfies $\ell$-diversity if, for each combination of quasi-identifier values in $\mathcal{QI}$, there are at least $\ell$ "well-represented" sensitive attribute values in the corresponding group of records.

\textbf{$t$-closeness}~\cite{4221659}:
    Given a dataset $\mathcal{D}$, a quasi-identifier $\mathcal{QI}$, and a parameter $t$, $\mathcal{D}$ satisfies $t$-closeness if, for each combination of quasi-identifier values in $\mathcal{QI}$, the distribution of sensitive attribute values in the corresponding group of records is "close" to the overall distribution of sensitive attribute values in the entire dataset.

\begin{table*}[t]
\caption{Studies on anonymization-based ILFPPM.}
\label{tab:Anon}
\centering
\scriptsize
\begin{tabular}{ccclclcl}
\toprule
\textbf{Paper}	& \textbf{Year}  & \textbf{Localization} & \textbf{Privacy Method} & \textbf{Dataset}  & \textbf{Metric} & \textbf{Adversary Model}  & \textbf{Attack Model}\\
\hline 

\cite{Anon-Kim2012} 
& 2012 & - 
& {\begin{tabular}{@{}l@{}} $k$-anonymity \\ Hierarchical graph  \end{tabular}}
& Simulated 
& {\begin{tabular}{@{}l@{}} Area of ASR \\ No. of cells \end{tabular}}
& -
& {\begin{tabular}{@{}l@{}} Loc. Priv. Attack \\ Inference Attack \end{tabular}}  \\ \hline 

\cite{Anon-Zhu2014} 
& 2014 
& kNN, SVM 
& {\begin{tabular}{@{}l@{}} $k$-anonymity \\ cloaking \\ obfuscation \end{tabular}}
& Simulated 
& {\begin{tabular}{@{}l@{}} Success Rate \\ Area of ASR \end{tabular}}
& - 
& Loc. Priv. Attack \\ \hline

\cite{Anon-Konstantinidis2015}
& 2015
& -
& {\begin{tabular}{@{}l@{}}  $k$-anonymity \\ Bloom Filter \end{tabular}} 
& {\begin{tabular}{@{}c@{}} Simulated \\ \hline CSUCY \\ KIOSUCY \\ Crawdad \end{tabular}} 
& {\begin{tabular}{@{}l@{}}  Energy Consumption \\ No. of Messages \\ Run Time \end{tabular}}
& Fully malicious 
& {\begin{tabular}{@{}l@{}}  Loc. \& Data Priv. Attack \\  Linking Attack \end{tabular}} \\ \hline

\cite{Anon-Kim2016}
& 2016
& kNN
& {\begin{tabular}{@{}l@{}} $k$-anonymity \\ $\ell$-diversity \\ Hierarchical graph \end{tabular}}
& Simulated 
& {\begin{tabular}{@{}l@{}}  Success Rate \\ Area of ASR  \\ Response Time \end{tabular}}
& Fully trusted 
& {\begin{tabular}{@{}l@{}}  Loc. Priv. Attack \\ Protocol-based Attack \\ Context Linking Attack \end{tabular}} \\ \hline

\cite{Anon-Alikhani2018}
& 2018
& NN
& {\begin{tabular}{@{}l@{}} Hilbert Curve  \\ $k$-anonymity \end{tabular}}
& Simulated 
& {\begin{tabular}{@{}l@{}} Coms. Overhead \\ Loc. Error\end{tabular}}
& - 
& - \\ \hline

\cite{Anon-Sazdar2020}
& 2020
& kNN
& {\begin{tabular}{@{}l@{}} $k$-anonymity \\ Randomization \\ Permutation \end{tabular}}
& {\begin{tabular}{@{}c@{}} Simulated \\ \hline CRI \end{tabular}} 
& {\begin{tabular}{@{}l@{}} Coms. Overhead \\ Loc. Error\end{tabular}}
& Fully malicious 
& {\begin{tabular}{@{}l@{}}  Loc. Priv. Attack \end{tabular}} \\ \hline

\cite{Anon-Zhao2020}
& 2020
& kNN
& $k$-anonymity
& Simulated
& {\begin{tabular}{@{}l@{}} Success Rate \\ CR Area \\ Coms. Overhead \\ Loc. Error\end{tabular}}
& -
& {\begin{tabular}{@{}l@{}} Loc. Priv. Attack \\ Context Linking Attack \end{tabular}} \\ \hline

\cite{Anon-Sazdar2021}
& 2021
& kNN
& {\begin{tabular}{@{}l@{}} $k$-anonymity \\ Hilbert Curve \\ Bloom Filter \end{tabular}} 
& {\begin{tabular}{@{}c@{}} Simulated \\ \hline CRI \\ CSUCY \\ KIOSUCY \\ PosData \end{tabular}}
& Loc. Error
& Semi-trusted 
& Loc. priv. Attack \\ \hline

\cite{FATHALIZADEH2022102665}
& 2022
& -
& {\begin{tabular}{@{}l@{}} $k$-anonymity \\ $\ell$-diversity \\ $t$-closeness \\ ($\alpha,k$)-anonymity \\ $\delta$-presence \end{tabular}} 
& {\begin{tabular}{@{}c@{}} Simulated \\ \hline CRI \\ UJIIndoorLoc \end{tabular}} 
& {\begin{tabular}{@{}l@{}}  Spatial Loss \\ RSS Loss \\ NCP \end{tabular}}
& Semi-trusted 
& {\begin{tabular}{@{}l@{}} Loc. \& Data Priv. Attack \\ Semantic Attack \\ Data Linking Attack \\ Probability-based Attack \end{tabular}} \\

\bottomrule
\end{tabular}
\end{table*}

\subsubsection{Comparative analysis of anonymization techniques}

Anonymization techniques in IPSs can protect individuals' privacy by masking identifiable information, while still allowing LBS to function. As depicted in Fig.~\ref{fig:k-anon}, these methods can be used for both location data (such as geographic coordinates) and trajectory data (movement over time). By anonymizing this data, IPSs lower the chances of re-identification, ensuring that sensitive information remains secure without compromising system performance.

Table~\ref{tab:Anon} reports the methods proposed for ILFPPMs based on anonymization. One of the earliest works by \cite{Anon-Kim2012} and \cite{Anon-Zhu2014} utilized $k$-anonymity combined with hierarchical graphs and cloaking mechanisms. These methods ensure that each individual is indistinguishable from at least $k$ other individuals within the dataset, mitigating the risk of re-identification. In these studies, performance is typically measured using metrics like Area of Anonymity Set Radius (ASR) and the number of cells, which indicate how effectively the location data is obfuscated. However, these methods may struggle to maintain high privacy guarantees against sophisticated adversaries, as they primarily target location privacy attacks and inference attacks, without addressing more advanced attack models like linking or semantic attacks.

In later studies, such as \cite{Anon-Konstantinidis2015}, $k$-anonymity is combined with Bloom filters to improve computational efficiency and scalability. This approach was evaluated using a variety of datasets, such as CSUCY and Crawdad, allowing the method to demonstrate its effectiveness in diverse environments. These methods showed improvements in terms of energy consumption, number of messages, and run time, which are critical in real-time indoor localization systems. However, the fully malicious adversary model adopted in this study highlights the vulnerability of these methods to more sophisticated attacks like linking and data privacy attacks. Therefore, while Bloom filters enhance efficiency, they require additional mechanisms to strengthen privacy guarantees in more hostile environments.

As privacy concerns grew, more advanced models like $\ell$-diversity and $t$-closeness emerged to address some of the inherent weaknesses of $k$-anonymity, particularly concerning attribute disclosure risks. For instance, [Kim 2016] integrated $\ell$-diversity into their $k$-anonymity framework, ensuring the diversity of sensitive attributes within each anonymity group is preserved. This model performed well against context-linking and protocol-based attacks while maintaining a high success rate and low response time. However, these methods typically assume a fully trusted adversary model, which may limit their effectiveness in real-world scenarios where such trust is unrealistic.

Recent works, such as \cite{FATHALIZADEH2022102665} and \cite{Anon-Sazdar2021}, have extended anonymization techniques by combining $k$-anonymity with spatial methods like Hilbert curves and randomization techniques. These studies highlight a growing trend toward improving the resilience of anonymization methods in dynamic environments with semi-trusted or fully malicious adversaries. The use of randomization and permutation in these works provides additional layers of security against location privacy attacks, though these methods often introduce higher communication overhead and location errors. Furthermore, \cite{FATHALIZADEH2022102665} adopts a comprehensive approach by incorporating multiple privacy models, such as $t$-closeness and $\delta$-presence, enabling protection against a wider range of attacks, including semantic and data linking attacks.


\subsubsection{Discussion on employing Anonymization-based methods}
To implement anonymization techniques in ILF, the following recommendations will help ensure both privacy protection and operational efficiency:

\medskip
\textbf{Selecting the right anonymity model:} 
Depending on the specific privacy needs and the strength of the adversary model, a range of anonymization techniques can be utilized in IPSs. One commonly used approach is $k$-anonymity, which ensures that an individual's data cannot be distinguished from at least $k-1$ others. However, $k$-anonymity may fall short against more sophisticated adversaries capable of performing inference attacks or cross-referencing external datasets. To address these vulnerabilities, enhanced models such as $\ell$-diversity or $t$-closeness can be implemented. $\ell$-diversity extends $k$-anonymity by ensuring that sensitive attributes within a group are diverse enough to prevent inference attacks, while $t$-closeness further strengthens protection by maintaining the distribution of sensitive data close to its overall population distribution, thereby mitigating risks of linking attacks and preventing adversaries from inferring sensitive information based on statistical imbalances. These extended techniques offer more robust privacy safeguards, especially in scenarios involving powerful adversaries or complex datasets.

\medskip
\textbf{Efficiency considerations:} 
The additional overhead brought by anonymization techniques, including the computational requirements of obfuscating data and the increased energy consumption, must be thoroughly assessed when designing IPSs. These costs can significantly impact the performance of the system, particularly in terms of processing speed, battery life in mobile devices, and the efficiency of data transmission. Some methods, such as Hilbert Curve anonymization, are specifically designed to minimize these overheads. By mapping spatial data onto a one-dimensional curve, Hilbert Curve anonymization reduces communication costs and decreases localization errors, making it highly effective for real-time applications that require both privacy protection and low-latency performance. This balance between privacy and system efficiency makes such methods ideal for deployment in resource-constrained environments, such as mobile devices or large-scale, real-time IPS networks. However, selecting the right anonymization technique requires careful consideration of trade-offs between privacy, computational overhead, and system responsiveness.


\medskip
\textbf{Trade-offs between privacy and utility:} 
The utility of anonymized data can be significantly diminished due to data suppression or generalization techniques employed to protect privacy. These modifications often degrade the accuracy of the data, making it less useful for applications like IPS or LBS. To maintain an effective balance between privacy and functionality, it is essential to evaluate metrics such as location error, which measures the deviation from actual position, and run time, which affects system performance. By analyzing these metrics, the optimal trade-off between safeguarding privacy and retaining the usability of the data can be determined. Techniques such as Hierarchical Graphs \cite{Anon-Kim2012} \cite{Anon-Kim2016} and Hilbert Curves \cite{Anon-Alikhani2018} \cite{Anon-Sazdar2021} are particularly effective in achieving this balance. Hierarchical Graphs allow for varying levels of data abstraction, which can be adjusted based on privacy needs, while still preserving important spatial relationships. Hilbert Curves, on the other hand, provide a space-filling curve that helps reduce location error while minimizing the impact of noise or suppression. Both methods offer superior trade-offs, ensuring higher location accuracy while simultaneously enhancing privacy, making them well-suited for privacy-preserving IPS where data utility is crucial.

\subsection{Differential Privacy in ILFPPM}
Here, we provide a brief explanation of the Differential Privacy (DP) mechanism and various DP-based solutions and subsequently discuss the DP-based ILFPPM. Finally, the guidelines on using DP for ILF are suggested.

\subsubsection{DP-based techniques}
\textbf{$\epsilon$-differential privacy}~\cite{10.1007/11681878_14}:
Consider a positive real number $\epsilon$ and a randomized algorithm denoted as $\mathcal{A}$, which takes a dataset as input, representing the actions of a trusted party holding the data. Let $\textit{im}(\mathcal{A})$ represent the set of possible outputs of the algorithm.
The algorithm $\mathcal{A}$ is considered to provide $\epsilon$-differential privacy if, for all pairs of datasets $\mathcal{D}$ and $\mathcal{D}'$ that differ in a single element (i.e., the data of one person), and for all subsets $S$ of $\textit{im}(\mathcal{A})$, the following inequality holds:

\begin{equation}
	\frac{\Pr[\mathcal{A}(\mathcal{D}) \in S]}{\Pr[\mathcal{A}(\mathcal{D}') \in S]} \leq e^\epsilon,
\end{equation}
where the probability is computed over the randomness used by the algorithm.

\begin{figure}
    \centering    \includegraphics[width=0.8\textwidth]{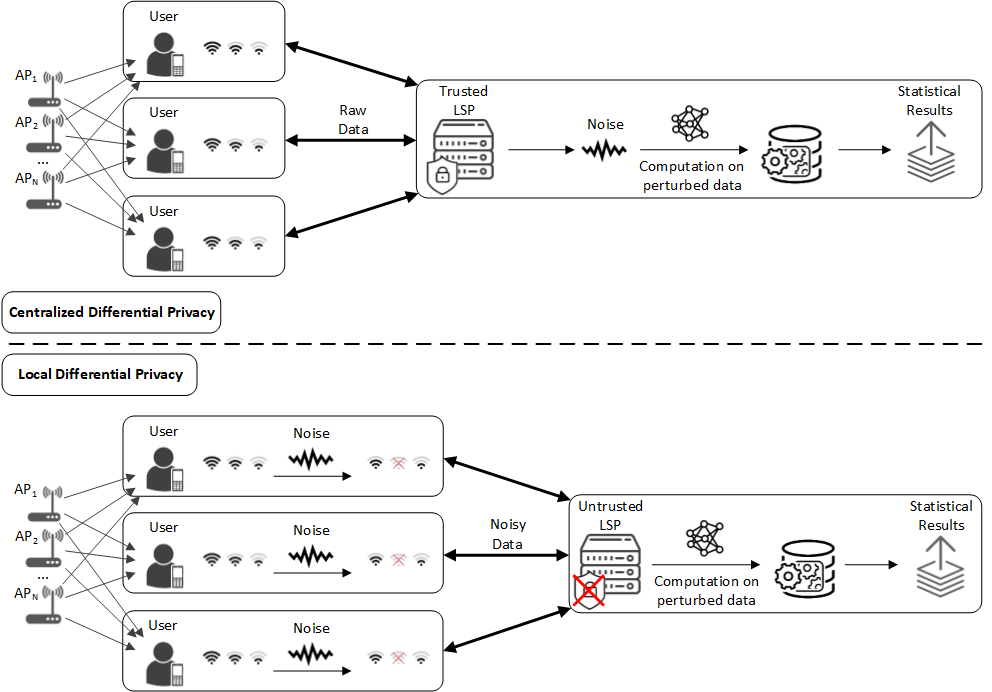}
    \caption{Centralized Differential Privacy vs. Local Differential Privacy in Indoor Positioning Systems}
    \label{fig:ldp_cdp}
\end{figure}

\medskip
\textbf{Geo-indistinguishability}~\cite{geo}:
When releasing aggregated information about a group of individuals, DP is an effective tool. However, it may not be suitable for scenarios involving a single individual, such as location privacy. DP prevents any change in an individual's location from affecting the published output, making it challenging to convey meaningful information to the service provider. 
To address these limitations, Geo-Indistinguishability was introduced. Geo-Indistinguishability adds random noise to a user's true location to prevent an adversary with background knowledge from accurately inferring the user's location. This approach utilizes 2D Euclidean distance, suitable for localization purposes, whereas previous methods primarily focused on Hamming distance. 
Formally, for a randomized mechanism $\mathcal{K}$ that reports an obfuscated location $z$ instead of the user's real location $x$, $\epsilon$-Geo-Indistinguishability is satisfied if, for any location ${x}'$:
\begin{equation}
	\mathcal{K}(x)(z) \leq e^{\epsilon d(x,{x}')} \cdot \mathcal{K}({x}')(z),
\end{equation}
where $d$ represents the Euclidean distance between $x$ and ${x}'$. Geo-Indistinguishability ensures that the probability of reporting point $z$ instead of other points $x$ and ${x}'$ differs by at most $e^{-\epsilon d(x,{x}')}$. This is achieved using a noise function, causing the probability of reporting a point around $z$ to decrease exponentially as the distance from the actual location $x$ increases. In a 1D linear space, this is accomplished using the Laplace distribution with the probability distribution function (PDF) $f_X(x) = \frac{\epsilon}{2}e^{-\epsilon \left |x-\mu  \right |}$, where $\mu$ and $\epsilon$ are parameters in the Laplace distribution.


\begin{table*}[t]
\caption{Studies on different DP-based ILFPPM.}
\label{tab:dp}
\centering
\scriptsize
\begin{tabular}{ccclclcl}
\toprule
\textbf{Paper}	& \textbf{Year}  & \textbf{Localization} & \textbf{Privacy Method} & \textbf{Dataset}  & \textbf{Metric} & \textbf{Adversary Model}  & \textbf{Attack Model}\\
\hline 

\cite{DP-Li2016}
& 2016
& kNN
& {\begin{tabular}{@{}l@{}} CDP \\ Paillier  \end{tabular}} 
& Private 
& {\begin{tabular}{@{}l@{}} Coms. Overhead \\ Loc. Error\end{tabular}}
& Semi-trusted 
& Loc. Priv. Attack \\ \hline 

\cite{DP-Sadhu2017}
& 2017
& -
& {\begin{tabular}{@{}l@{}} LDP \\ Two-step classiﬁer  \end{tabular}}  
& {\begin{tabular}{@{}c@{}} Multi Building \\ Rutgers Uni \end{tabular}} 
& Loc. Error
& - 
& - \\ \hline 

\cite{DP-Zhu2017}
& 2017
& -
& {\begin{tabular}{@{}l@{}} LDP \\ Delaunay triangulation \\ Range generalization \end{tabular}} 
& Simulated 
& Loc. Error
& - 
& - \\ \hline 

\cite{DP-WookKim2018}
& 2018
& -
& {\begin{tabular}{@{}l@{}} LDP \\ Randomized Response \end{tabular}} 
& {\begin{tabular}{@{}c@{}} Simulated \\ \hline Sangmyung Uni \end{tabular}} 
& Error Rate
& Fully malicious
& {\begin{tabular}{@{}l@{}} Loc. Priv. Attack \\ Probability-based Attack  \end{tabular}} \\ \hline 

\cite{DP-Wang2018}
& 2018
& -
& {\begin{tabular}{@{}l@{}} CDP \\ AP fuzzification \\ Finger Clustering \\ Finger Permutation \end{tabular}}  
& PosData
& {\begin{tabular}{@{}l@{}} Loc. Error \\ Distance Error\end{tabular}}
& - 
& {\begin{tabular}{@{}l@{}} Loc. \& Data Priv. Attack \\ Probability-based Attack  \end{tabular}} \\ \hline 

\cite{DP-Zhao2018}
& 2018
& -
& {\begin{tabular}{@{}l@{}} LDP \\ Segmenting Data \\ $k$-anonymity \end{tabular}}   
& Private
& {\begin{tabular}{@{}l@{}} Coms. Overhead \\ Loc. Error\end{tabular}}
& Semi-trusted
& {\begin{tabular}{@{}l@{}} Loc. Priv. Attack \\ Context linking Attack \\ Probability-based Attack \\ Protocol-based Attack  \end{tabular}} \\ \hline 

\cite{DP-Kim2019}
& 2019
& -
& {\begin{tabular}{@{}l@{}} LDP \\ Randomized Encoding \end{tabular}}   
& {\begin{tabular}{@{}c@{}} Simulated \\ \hline Beijing Taxi \\ Trajectory \end{tabular}}
& Loc. Error
& - 
& - \\ \hline

\cite{DP-Navidan2022}
& 2022
& -
& {\begin{tabular}{@{}l@{}} LDP \\  Local Hashing \\ Unary Encoding \\ Histogram Encoding \\ Random response \\ RAPPOR \end{tabular}}   
& {\begin{tabular}{@{}c@{}} JUIndoorLoc \\ CRI \end{tabular}}
& Loc. Error
& - 
& Loc. \& Data Priv. Attack \\ \hline

\cite{DP-Zhang2022}
& 2022
& FSELM
& {\begin{tabular}{@{}l@{}} LDP \\ Edge–cloud \\ collaboration \end{tabular}}  
& Private
& {\begin{tabular}{@{}l@{}} Coms. Overhead \\  Loc. Error \\ Run Time \end{tabular}}
& Fully malicious 
& {\begin{tabular}{@{}l@{}} Loc. Priv. Attack \\ Probability-based Attack \\ Inference Attack \end{tabular}} \\ \hline

\cite{DP-Min2022}
& 2022
& -
& {\begin{tabular}{@{}l@{}} GeoInd \\ Laplace Mechanism \end{tabular}} 
& Simulated
& {\begin{tabular}{@{}l@{}} Coms. Overhead \\  Loc. Privacy \\ QoS Loss \end{tabular}}
& Fully malicious
& {\begin{tabular}{@{}l@{}} Loc. Priv. Attack \\ Probability-based Attack \\ Inference Attack \end{tabular}}  \\ \hline

\cite{DP-Ftl2023}
& 2023
& -
& {\begin{tabular}{@{}l@{}} GeoInd \\ Laplace Mechanism \\ Distance calculation \\ RSS Generation \end{tabular}}
& {\begin{tabular}{@{}c@{}} Simulated \\ \hline JUIndoorLoc \\ CRI \end{tabular}} 
& {\begin{tabular}{@{}l@{}}  Loc. Error \\ Probabilistic \\ QoS Loss \end{tabular}}
& Fully malicious
& {\begin{tabular}{@{}l@{}} Loc. Priv. Attack \\ Probability-based Attack \\ Context Linking Attack \end{tabular}}  \\ \hline

\cite{DP-Min2023}
& 2023
& -
&{\begin{tabular}{@{}l@{}} GeoInd \\ Reinforcement learning \end{tabular}}
& Simulated 
& {\begin{tabular}{@{}l@{}} Coms. Overhead \\  Loc. Privacy \\ QoS Loss \end{tabular}}
& Fully malicious
& {\begin{tabular}{@{}l@{}} Inference Attack \\ Semantic Attack \\ Probability-based Attack \end{tabular}}   \\ \hline

\cite{inference_2024}
& 2024
& -
&{\begin{tabular}{@{}l@{}} LDP \\ GeoInd \end{tabular}}
& {\begin{tabular}{@{}c@{}} Geolife \\ Gowalla \end{tabular}} 
&  QoS Loss
& Fully malicious
& {\begin{tabular}{@{}l@{}} Inference Attack \\ Loc. Priv. Attack \\ Probability-based Attack \end{tabular}}   \\

\bottomrule
\end{tabular}
\end{table*}

\subsubsection{Comparative analysis of DP-based techniques}
The privacy norm of DP is also utilized to protect privacy in indoor localization. The proposed methods lie on one of the Centralized Differential Privacy (CDP), Local Differential Privacy (LDP), or Geo-indistinguishability.
CDP is a framework for protecting the privacy of individual data points in a centralized dataset. In this approach, a trusted server applies DP mechanisms to the entire dataset to ensure that the analysis or queries performed on the data do not reveal sensitive information about any individual data point. In LDP, on the other hand, individual data contributors apply privacy mechanisms to their data before sending it to a central server or aggregator. Each data point is perturbed or obfuscated in such a way that, when aggregated with other data points, the perturbed data still provides statistically valid information for analysis, but does not compromise the privacy of the individuals. This difference between CDP and LDP is shown in Fig.~\ref{fig:ldp_cdp}. In addition, geo-indistinguishability is a DP-based technique used to protect the privacy of individuals' locations by adding noise to local data, thus leading to obfuscating the real location.

Table~\ref{tab:dp} provides a list of ILFPPM based on DP.
One of the earliest implementations of DP-based ILFPPM is presented by \cite{DP-Li2016}, which integrates CDP with Paillier encryption for protecting kNN-based localization. This method focuses on communication overhead and location error as performance metrics, showing its potential in a semi-trusted adversary model. However, location privacy attacks remain a critical vulnerability, indicating the need for more refined adversary models to address advanced threats.
In contrast, \cite{DP-Sadhu2017} and \cite{DP-Zhu2017} employed LDP mechanisms, offering a more decentralized privacy guarantee. \cite{DP-Sadhu2017}’s work applies two-step classification for building-level indoor localization using real-world datasets such as Rutgers University. \cite{DP-Zhu2017}, on the other hand, adopts Delaunay triangulation and range generalization, focusing heavily on reducing location error. Although both approaches show promise, they lack explicit adversary models, which makes it difficult to gauge their robustness against more sophisticated privacy attacks.

The aforementioned works concentrate on privacy preservation within specific types of indoor localization. 
The work described in~\cite{DP-Zhao2018} introduces a privacy-preserving paradigm-driven framework for indoor localization utilizing LDP. This approach is motivated by the observation that many IPS adhere to a common two-stage localization paradigm. 
As LDP evolved, \cite{DP-WookKim2018} introduced randomized response as an LDP mechanism, showing improvements in reducing error rates under fully malicious adversaries. This study further demonstrated the method’s efficacy in countering location privacy and probability-based attacks using Sangmyung University data. However, despite its strengths, randomized response techniques often introduce noise, potentially increasing the communication overhead or compromising data utility.
Moreover, the study presented in~\cite{DP-Kim2019} combines local differential privacy and optimal data encoding to disturb users' data. 
Similarly, authors in~\cite{DP-Wang2018} propose a DP-based privacy-preserving indoor localization scheme that introduces noise to the data, such as RSSI, CSI, and so forth, which increases the likelihood of significant localization errors and lowers the standard of localization services. 

The work in~\cite{DP-Zhang2022} applies fusion of signal gathered from multiple wireless technologies (e.g., WiFi and BLE) with differentially private fingerprint fusion semi-supervised extreme learning machine for indoor localization in the edge computing, called Adp-FSELM. It incorporates edge-cloud collaboration to optimize communication overhead and run time, while maintaining a balance between location error and privacy protection in the presence of fully malicious adversaries.
Another study in ~\cite{DP-Navidan2022} introduces a novel privacy-aware framework for aggregating indoor location data employing the LDP technique. It combines multiple privacy-preserving techniques like local hashing, unary encoding, and RAPPOR, using datasets like JUIndoorLoc and CRI. In this work, the user location data is locally transformed on the user's device and subsequently sent to the aggregator. This comprehensive approach is particularly effective against location and data privacy attacks, showcasing the versatility of LDP in practical scenarios.

Employing other DP-based solutions, the authors in~\cite{DP-Min2022} initially introduce 3D Geo-indistinguishability for the application of indoor localization. This is followed by the addition of reinforcement learning for the sake of semantic location privacy in~\cite{DP-Min2023}. These papers assume that the user has access to her location and sends her location for obtaining services. From another perspective,~\cite{DP-Ftl2023} introduces two methods for applying geo-indistinguishability based on RSS vectors without having access to the location.
These studies strongly emphasize balancing location privacy and QoS loss while addressing increasingly complex attack models like semantic, inference, and context-linking attacks. The trade-off between privacy guarantees and QoS is a recurring challenge, indicating the need for further refinement of DP-based ILFPPM to meet real-world demands without sacrificing data utility.

\subsubsection{Discussion on employing DP-based techniques}
To implement DP-based techniques in indoor localization systems, the following guidelines will help ensure both privacy protection and operational efficiency:

\medskip
\textbf{Selection of DP model:} 
When choosing between CDP and LDP, trust assumptions play a key role. CDP is suitable when a trusted centralized authority is available, as it adds noise to aggregated data at the server level, allowing for global optimization. However, in cases where users don’t fully trust the data aggregator or want to minimize reliance on a central entity, LDP is more appropriate. LDP ensures privacy at the user level by adding noise before data leaves the device, making it ideal for decentralized environments or scenarios where user control and anonymity are crucial. The choice depends on balancing privacy, trust, and system architecture.

\medskip
\textbf{Mitigating attack models:} 
In environments vulnerable to fully malicious adversaries, such as those discussed by LDP and CDP, more advanced privacy mechanisms like randomized response, edge-cloud collaboration, or GeoInd are often necessary. These methods provide stronger resilience against a broader spectrum of attacks, including location privacy breaches, probabilistic attacks, and sophisticated inference-based threats. Randomized response allows individual users to inject randomness into their data, supporting privacy at the source level. GeoInd \cite{DP-Ftl2023}, by introducing geographic obfuscation, further shields user location data, making it harder for adversaries to infer sensitive information. 

\medskip
\textbf{Balancing privacy and utility:} 
Implementations must achieve a careful balance between privacy protection and data utility. LDP techniques commonly inject noise to enhance privacy, which can result in a reduction in data utility, often evaluated using metrics such as location error or QoS loss. It is crucial to select mechanisms that deliver the required level of privacy while ensuring that the noise introduced remains within acceptable limits for the localization accuracy. This balance allows for maintaining effective data usability and performance, enabling systems to function optimally without compromising user privacy. Additionally, evaluating various LDP methods can help identify those that minimize adverse impacts on utility, facilitating better overall system performance and user satisfaction.

\medskip
\textbf{Optimization of communication overhead:} 
When designing real-time systems or large-scale deployments, it is vital to assess the communication overhead associated with the selected DP mechanism. Techniques like randomized encoding \cite{DP-Kim2019} and Laplace mechanisms \cite{DP-Ftl2023} can introduce significant overhead, potentially impacting overall system performance and responsiveness. This added complexity may lead to slower data processing and increased latency, which are critical in applications requiring immediate feedback or high throughput. Therefore, carefully evaluating the trade-offs between privacy protection and operational efficiency is essential, especially in resource-constrained environments where CPU power, bandwidth, and energy are limited. By considering these trade-offs, developers can select a DP approach that safeguards user privacy while maintaining system functionality and responsiveness, ensuring optimal performance across various operational contexts.


\subsection{Federated Learning in ILFPPM}
Here, we briefly explain the FL mechanism and then discuss the existing FL-based ILFPPM. Finally, we provide guidelines on employing FL in ILF systems.

\subsubsection{Federated learning mechanism}
\begin{figure} [t]
    \centering
    \includegraphics[width=0.8\textwidth]{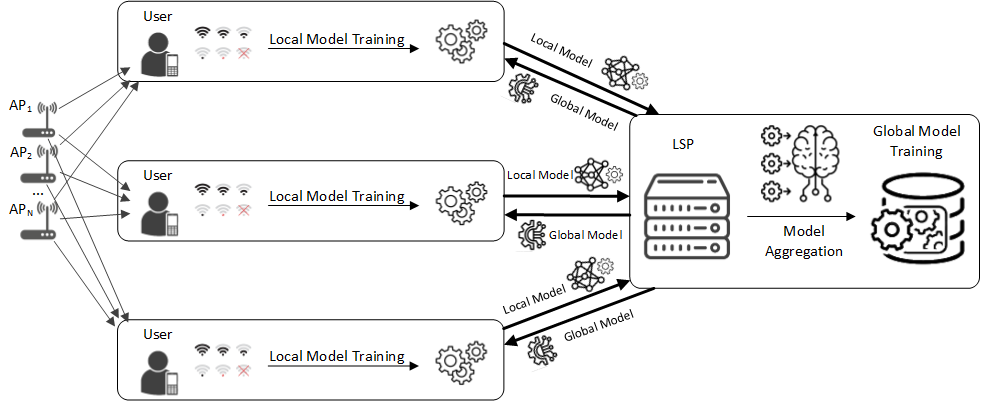}
    \caption{An overview of FL in ILF systems}
    \label{fig:fl}
\end{figure}

Google introduced FL, a privacy-preserving distributed ML approach, in 2016~\cite{mcmahan2017communication}. This method trains an algorithm on several decentralized edge devices or local servers without requiring them to exchange raw data, in contrast to traditional centralized machine learning techniques that load all local datasets onto a single server. Initially, a machine learning model is trained by a centralized server and this model is then sent to local devices that hold their own private data. On these respective devices, the local models are trained using their private data, as shown in Fig.~\ref{fig:fl}. The training process typically involves using local computational resources and updates the model based on the local data. After training, the locally updated models send their changes (e.g., parameters or gradients) back to the centralized server, which aggregates these updates. Finally, the server updates the global model using an aggregation method such as FedAvg and FedAmp \cite{hsu2019measuring},\cite{wu2021personalized}. FedAvg is one of the most common aggregation methods for FL, where the central server combines the gradients sent by the users using the below equation and updates the global model.
\begin{equation}
w_{t+1}  \leftarrow  \sum_{k=1}^{K}\frac{n_{k}}{n} w^{k}_{t}, 
\end{equation}
where $n$ is the total number of data, $n_{k}$ is the number of samples for the $k$’th user, $w_{t+1}$ is the gradients of global model in the round $t+1$, $w^{k}_{t}$ is the gradients of $k$'th user's local model in the round $t$. FL is especially useful in scenarios where data privacy or data locality is a concern. It allows for collaborative model training without the need to centralize sensitive data, thus mitigating privacy and security risks. By leveraging FL, systems can benefit from the collective intelligence of decentralized data sources while protecting user privacy and regulations.

\begin{table*}[t]
\caption{Studies on FL-based ILFPPM.}
\label{tab:fl}
\centering
\scriptsize
\begin{tabular}{ccclcll}
\toprule
\textbf{Paper}	& \textbf{Year}  & \textbf{Localization} & \textbf{Privacy / Aggregation Method} & \textbf{Dataset}  & \textbf{Metric}  & \textbf{Attack Model}\\
\hline 

\cite{FL-Liu2019}
& 2019
& DNN
& AutoEncoder
& Private 
& {\begin{tabular}{@{}l@{}} Coms. Overhead \\  Loc. Error \end{tabular}}
& Loc. Priv. Attack   \\  \hline

\cite{FL-Li2020}
& 2020
& SAEC
& FedAvg
& UJIIndoorLoc 
& Loc. Error
& Loc. Priv. Attack   \\  \hline

\cite{FL-Ciftler2020}
& 2020
& MLP
& FedAvg
& UJIIndoorLoc 
& Loc. Error
& -   \\  \hline

\cite{FL-PWu2021}
& 2021
& MLP
& {\begin{tabular}{@{}l@{}}  FedAvg \\ FedAmp \\ FedAmp-fusion \end{tabular}}
& UJIIndoorLoc 
& Loc. Error
& -   \\  \hline

\cite{FL-Wu2021}
& 2021
& MLP
& {\begin{tabular}{@{}l@{}} FedAvg \\ DP \end{tabular}}
& {\begin{tabular}{@{}c@{}}  UJIIndoorLoc \\ Guangzhou Xinguang \\ shopping mall \end{tabular}} 
& {\begin{tabular}{@{}l@{}}  Loc. Error \\ Noise Effect \end{tabular}}
& {\begin{tabular}{@{}l@{}}  Inference Attack \\ Loc. Priv. Attack \end{tabular}}    \\  \hline

\cite{FL-Tasbaz2022}
& 2022
& {\begin{tabular}{@{}l@{}} kNN, MLP \\ Random Forest \end{tabular}} 
& FedAvg
& {\begin{tabular}{@{}c@{}}  UJIIndoorLoc \\ CRI \end{tabular}} 
& {\begin{tabular}{@{}l@{}} Accuracy \\ Run Time \end{tabular}}
& {\begin{tabular}{@{}l@{}} Loc. Priv. Attack \\ Probability-based Attack \end{tabular}}     \\  \hline

\cite{FL-Cheng2022}
& 2022
& MLP 
& FedAvg
& UJIIndoorLoc       
& {\begin{tabular}{@{}l@{}} Accuracy \\  Loc. Error \end{tabular}}
& -  \\  \hline

\cite{FL-Wu2022}
& 2022
& MLP 
& {\begin{tabular}{@{}l@{}}  FedAvg \\ Knowledge distillation \\ Mixture of Experts \end{tabular}}
& {\begin{tabular}{@{}c@{}}  UJIIndoorLoc \\ Guangzhou Xinguang \\ shopping mall \end{tabular}}       
& Loc. Error
& {\begin{tabular}{@{}l@{}}  Inference Attack \\ Loc. Priv. Attack \end{tabular}}   \\  \hline

\cite{FL-Nagia2022}
& 2022
& CNN, LSTM 
& FedProx
& Private       
& Loc. Error
& -  \\  \hline

\cite{FL-dropout-2022}
& 2022
&  NN
&  Monte Carlo Dropout
& UJIIndoorLoc       
& {\begin{tabular}{@{}l@{}} Loc. Error \\ Time Cost \end{tabular}}
& - \\ \hline

\cite{FL-Graph-2022}
& 2022
& GNN
& Federated Graph Learning
& {\begin{tabular}{@{}c@{}} Guangzhou Xinguang \\ shopping mall \end{tabular}}       
& Loc. Error 
& - \\ \hline

\cite{FL-Gao2023}
& 2023
& CNN, DNN 
& AutoEncoder
& UJIIndoorLoc       
& {\begin{tabular}{@{}l@{}}  Accuracy \\ Loc. Error \\ Coms. Overhead \end{tabular}}
& {\begin{tabular}{@{}l@{}}  Inference Attack \\ Loc. Priv. Attack \end{tabular}}   \\  \hline

\cite{FL-Kumar2023}
& 2023
& CNN, LSTM 
& -
& Private       
& Accuracy
& -  \\  \hline

\cite{FL-Gufran2023}
& 2023
& MLP
& {\begin{tabular}{@{}l@{}} AutoEncoder \\  FedHIL \end{tabular}}
& Private       
& {\begin{tabular}{@{}l@{}} Coms. Overhead \\  Loc. Error \end{tabular}}
& Inference Attack \\  \hline

\cite{FL-Guo2023}
& 2023
& CNN 
& {\begin{tabular}{@{}l@{}} Cloud Feature Extractor \\ Homomorphic Encryption \end{tabular}}
& Simulated       
& {\begin{tabular}{@{}l@{}} Coms. Overhead \\  Loc. Error \end{tabular}} 
& {\begin{tabular}{@{}l@{}}  Inference Attack \\ Probability-based Attack \end{tabular}} \\ \hline

\cite{FL-kNN-2023}
& 2023
&  kNN
& {\begin{tabular}{@{}l@{}} Discrete Coordinate Encryption \\ Federated KNN \end{tabular}}
& Private       
&  {\begin{tabular}{@{}l@{}} Loc. Error \\ Energy Consumption \end{tabular}}
& - \\ \hline

\cite{FL-CSI-2024}
& 2024
&  CNN
&  Inner Production
&  Private      
&  {\begin{tabular}{@{}l@{}} Accuracy \\ Loc. Error \end{tabular}}
& Data Priv. Attack \\ \hline

\cite{FL-Etibadi2022}
& 2024
& DNN 
& {\begin{tabular}{@{}l@{}}  Knowledge distillation \\ Federated distillation \end{tabular}}
& {\begin{tabular}{@{}c@{}} Simulated \\ \hline UJIIndoorLoc \end{tabular}}       
& {\begin{tabular}{@{}l@{}} Coms. Overhead \\  Loc. Error \end{tabular}}
& -  \\  \hline

\cite{FL-Yan2024}
& 2024
& CNN
& FedAvg
& Private
&  {\begin{tabular}{@{}l@{}} Accuracy \\ Loc. Error \end{tabular}}
& - \\ \hline

\cite{FL-Tasbaz2024}
& 2024
& MLP, LSTM
& {\begin{tabular}{@{}l@{}} FedAvg \\  FedSGD \end{tabular}}
& SPAWC
& {\begin{tabular}{@{}l@{}} Coms. Overhead \\  Loc. Error \end{tabular}}
& Inference Attack \\ \hline

\cite{FL-Etibadi2024}
& 2024
& {\begin{tabular}{@{}l@{}} kNN \\ SVM  \end{tabular}} 
& {\begin{tabular}{@{}l@{}}  Transfer Learning \\ Meta-learning \end{tabular}}
& {\begin{tabular}{@{}c@{}} UJIIndoorLoc \\ Tampere Uni \\ Uni of Minho \end{tabular}}       
& {\begin{tabular}{@{}l@{}} Loc. Error \\ Adaptation speed \end{tabular}}
& -  \\  

\bottomrule
\end{tabular}
\end{table*}

\subsubsection{Comparative analysis of FL-Based techniques}
As FL enables training a centralized model across decentralized devices or servers holding local data samples, without exchanging the data samples, it has been applied to IPS. IPS often involves collecting location data from various devices and using FL allows these devices to collaboratively learn a positioning model without sharing the raw location data. Accordingly, instead of sending all raw data to the central server, only model updates or aggregated information need to be communicated, thus reducing the amount of data transmitted over the network. Since indoor locations vary very widely, this approach is very useful when updating the model. So under the FL setting, indoor positioning models can be trained using data from various users in different environments without having to centralize that data. This helps protect the privacy of users while still allowing for model improvements and leads to a more accurate and robust IPS. 


Table~\ref{tab:fl} reports ILFPPM studies in the FL setting. Most studies focus on the RSS data, but CSI is also taken into consideration in some papers~\cite{FL-Guo2023}. 
The reviewed studies employ AutoEncoders \cite{FL-Liu2019}, \cite{FL-Gao2023}, \cite{FL-Gufran2023} in order to have the dimensionality reduction and privacy preserving data sharing when users can share encoded data.
The most used aggregation method is also Federated Averaging (FedAvg) \cite{FL-Li2020}, \cite{FL-Ciftler2020}, \cite{FL-Wu2022}. More advanced approaches, such as Federated Proximal (FedProx) \cite{FL-Nagia2022}, Federated Graph Learning \cite{FL-Graph-2022}, and Federated Meta-Learning \cite{FL-Etibadi2024}, have been explored for specialized tasks. Additional methods, such as Monte Carlo Dropout \cite{FL-dropout-2022} and Federated Knowledge Distillation \cite{FL-Etibadi2022}, \cite{FL-Wu2022}, represent further advancements within the FL paradigm. These methods emphasize the increasing diversity of privacy-enhancing techniques specified to different attack models. 

In terms of datasets and performance metrics, the UJIIndoorLoc dataset is the most frequently used for evaluating these techniques, providing a consistent basis for measuring localization error and communication overhead~\cite{FL-Li2020}, \cite{FL-Tasbaz2022}, \cite{FL-Gao2023}. Private datasets \cite{FL-Liu2019}, \cite{FL-Nagia2022}, \cite{FL-Guo2023} have also been utilized to test performance under various privacy models. Common metrics include localization error, communication overhead, accuracy, runtime, and energy consumption. Some studies also account for noise effects \cite{FL-Wu2021} and time cost \cite{FL-dropout-2022}, reflecting a growing focus on the practical deployment and efficiency of these systems.

Many studies concentrate on defending against location privacy attacks and inference attacks \cite{FL-Wu2022}, \cite{FL-Gao2023}, \cite{FL-Guo2023}. Additionally, some works address probability-based and data privacy attacks \cite{FL-Tasbaz2022}, \cite{FL-CSI-2024}. However, certain studies do not specify a particular attack model \cite{FL-Ciftler2020}, \cite{FL-Nagia2022}, which may limit the general applicability of their privacy-preserving techniques.

There is a clear trend toward enhancing FL models by incorporating more sophisticated techniques. Earlier works utilized AutoEncoders with traditional privacy methods~\cite{FL-Liu2019}, while recent studies have integrated homomorphic encryption, federated knowledge distillation, and a mixture of experts to improve privacy without sacrificing performance \cite{FL-Wu2022}, \cite{FL-Guo2023}. The growing use of deep learning models, such as CNNs, LSTMs, and GNNs, highlights a trend toward employing more complex architectures to achieve better performance and privacy preservation.

\subsubsection{Discussion on employing existing FL-based techniques}
To apply FL in ILF systems, the following recommendations can help maintain privacy while optimizing system performance:

\medskip
\textbf{Selecting the appropriate aggregation and privacy methods:} 
When the primary concern is protecting against location privacy attacks, combining FedAvg with AutoEncoders offers an effective trade-off between security and computational efficiency. This approach ensures a reasonable level of privacy while maintaining system performance. However, for more complex threats, such as inference attacks or data privacy breaches, more sophisticated techniques may be necessary. Incorporating homomorphic encryption or federated knowledge distillation can significantly enhance the system's resilience to such attacks, albeit at the cost of increased computational overhead. These advanced methods offer stronger privacy guarantees, particularly in scenarios where sensitive data is more vulnerable to sophisticated adversarial models.


\medskip
\textbf{Balancing performance and privacy:} 
When performance metrics such as runtime and communication overhead are critical considerations, it is important to focus on methods that specifically address these concerns. Approaches like Monte Carlo Dropout and FedAvg are well-suited for maintaining a balance between high localization accuracy and reduced computational demands. By employing these techniques, it is possible to optimize system performance without significantly compromising on privacy or accuracy. These methods are particularly valuable in real-time applications, where minimizing time costs and communication overhead is essential for efficient system deployment and user experience.

\medskip
\textbf{Handling adversarial models:} 
When a system is exposed to multiple types of attack models, including inference attacks or probability-based attacks, advanced privacy-preserving techniques become essential. Methods such as Federated Knowledge Distillation and homomorphic encryption are particularly well-suited to these scenarios, as they are designed to safeguard sensitive data against more sophisticated and adaptive adversaries. These techniques provide strong defenses while ensuring that system functionality and performance are not significantly compromised. By integrating these advanced methods, the system can maintain high levels of privacy and security, even in the face of evolving and complex threats, without sacrificing its overall utility or user experience.

\subsection{Overall Comparison}
{In this section, first, we highlight key insights in ILF privacy preservation. Second, we explain a comparative assessment of ILFPPM techniques.
Finally, we offer practical guidelines for ILFPPM deployment. }

\subsubsection{Key observations}
The trade-off between privacy and performance in IPS highlights a significant challenge. Cryptographic techniques provide the strongest privacy guarantees but need high computational and communication overhead. In contrast, DP and FL offer a more practical balance between privacy and performance, though they may introduce localization errors or degrade QoS, reflecting the difficulty in optimizing both privacy and system efficiency.

Most privacy-preserving methods focus on location privacy attacks, while few address advanced attacks such as inference or data privacy attacks. Additionally, there is limited attention to fully malicious adversaries, especially within FL settings, indicating the need for more robust solutions to defend against sophisticated attacks.

The UJIIndoorLoc dataset is the most commonly used across all methods, serving as a standard benchmark for performance comparisons. However, this reliance on a single dataset limits the generalizability of solutions to diverse real-world scenarios. Cryptographic techniques, in particular, are often evaluated using private datasets or simulated environments, narrowing their applicability. Broader testing across varied datasets would improve the relevance and adaptability of these methods.

\subsubsection{Comparative assessment of ILFPPM}

{\fontsize{7.2}{8.7}\selectfont
\begin{table*}[t]
\caption{ {Comparative assessment of ILFPPM.}}
\label{tab:ilppm}
\centering
\small
\begin{tabular}{l llll}
\toprule
\textbf{ILFPPM}	\,\,& \textbf{Device/Transmission/Server}    \\ 
\midrule 
Cryptography & Transmission \\ 
Anonymization & Device/Server  \\ 
Differential Privacy & Device/Server  \\
Federated Learning & Device  \\
\bottomrule
\end{tabular}
\end{table*}
}


Cryptography-based ILFPPM primarily focuses on protecting the privacy of the data being processed, rather than securing the communication channel itself. While secure communication channels, such as those protected by Transport Layer Security (TLS), safeguard against external adversaries, cryptographic protocols in privacy-preserving indoor localization methods operate on top of these secure channels. The primary goal of these cryptographic protocols is to ensure that the data remains private even if the party at the other end of the channel is potentially malicious. However, anonymization-based solutions implement lightweight protocols either on the device or server to obfuscate the localization mechanism. DP-based approaches encompass various solutions, including CDP, LDP, and geo-indistinguishability each addressing different aspects of the privacy norm within DP. Depending on the context, DP can be applied to both the server and the user. FL-based solutions aim to decentralize the learning mechanism for users to preserve privacy. 
As shown in Table~\ref{tab:ilppm}, each privacy-preserving method, such as cryptographic methods employed primarily during transmission, anonymization implemented on both servers and devices, DP utilized across devices and servers, and FL mainly applied on devices. 

\subsubsection{Comprehensive guidelines for employing existing ILFPPM}
Comprehensive guidelines for utilizing existing ILFPPM help ensure strong privacy safeguards while maintaining system efficiency:

\medskip
\textbf{1) Choosing the right privacy method:}
If location privacy is a primary concern and the threat comes from semi-trusted adversaries, LDP presents an ideal solution. LDP strikes a strong balance between privacy and performance by ensuring sensitive data remains protected while minimizing computational overhead, making it well-suited for environments where both privacy and efficiency are critical.
For situations with extremely sensitive data where protection against fully malicious adversaries is essential, cryptographic techniques such as Homomorphic encryption or Paillier Encryption should be considered. These methods offer robust privacy guarantees by allowing computations on encrypted data without exposing sensitive information. Although resource-intensive, they ensure that data remains secure even in highly adversarial conditions.
In cases where privacy must be preserved without transmitting raw location data, FL-based approaches are highly effective. These methods facilitate decentralized data processing, making them advantageous for distributed systems, such as mobile applications or IoT-based indoor localization systems. FL enables collaborative learning across multiple devices without sharing sensitive information, reducing the risk of exposing location data while allowing accurate model training and system functionality.

\medskip
\textbf{2) Balancing privacy and performance:}
When real-time performance is a critical factor, it is advisable to adopt privacy methods that reduce both communication and computational overhead. Techniques such as anonymization and Lightweight LDP tend to be more efficient in these contexts, although they might entail a trade-off by sacrificing some degree of privacy protection. 
In cases where computational resources and energy consumption are constrained, which is more common in mobile or IoT devices, it is essential to select methods specifically designed for efficiency. Approaches like Federated Knowledge Distillation and AutoEncoders are well-suited for these environments, as they optimize resource usage while still maintaining reasonable levels of privacy. These methods allow for effective data processing and model training without overburdening limited computational capacities, enabling seamless operation in resource-constrained settings.

\medskip
\textbf{3) Handling different adversary models:}
To effectively defend against inference attacks or context-linking attacks, utilizing advanced cryptographic solutions such as Homomorphic encryption or to combine FL with DP are considered. These methods offer robust features that can significantly enhance privacy protection in indoor environments where sensitive data is at risk of exposure.
For lightweight systems that need to maintain strong privacy against location-based attacks, techniques such as randomized encoding and random response mechanisms are excellent options. These methods provide sufficient protection while incurring minimal performance costs, making them suitable for applications with limited computational resources. 

\medskip
\textbf{4) Deploying in real-world systems:}
FL techniques have demonstrated significant potential for application in real-world environments, such as smart homes, indoor navigation systems, and industrial IoT networks. By distributing computational tasks across multiple devices, these methods effectively reduce the risks associated with data transfer, enhancing both privacy and security. 
When implementing FL, it is essential to customize the approach to meet the specific requirements of the system in question. For instance, using FedAvg is ideal for scenarios where the data distribution is relatively balanced across participating devices. In contrast, if the data is non-independent and identically distributed (non-IID), which is the case in ILF, utilizing methods like FedProx can help address the challenges posed by this type of data variability. By carefully selecting and adapting FL techniques, we can optimize performance, maintain privacy, and ensure that their systems are robust and efficient in real-world applications. 


\section{Indoor Localization Applications and Privacy Concerns} \label{sec:app}

In recent years, indoor localization advances and the proliferation of mobile devices have made LBS more accessible. The applications of indoor localization include but are not limited to:

\subsection{Marketing}
Contextually-aware location-based marketing strategies have the potential to significantly boost sales and profits in e-commerce by delivering tailored marketing directly to consumers based on their location, especially in shopping centers, utilizing location-based technology. This approach allows real-time communication between consumers and sellers, enhancing the overall shopping experience. Additionally, it aids businesses in tracking customer behaviors, patterns, and foot traffic, offering personalized services based on mobile devices' information~\cite{MLSurvey-nessa, MLSurvey-navit}. However, these strategies bring privacy challenges, as the use of location-based technology for personalized marketing raises concerns about user privacy and potential misuse of real-time location data. Tracking customer behaviors without explicit consent and the extensive collection of personal data through personalized mobile devices also pose ethical and privacy issues. Furthermore, the precision of IPS in localizing mobile devices indoors and outdoors intensifies concerns about intrusive tracking if not appropriately safeguarded.

\subsection{Ambient Assisted Living and Disaster Management}
Ambient assisted living platforms, relying on precise indoor location tracking through technologies like Bluetooth and various IPS, offer significant benefits for the elderly, ill, or disabled individuals, particularly those with neurodegenerative diseases~\cite{7194823}. These platforms facilitate behavioral tracking, monitoring daily activities, movement patterns, vital signs, and detecting potential dangers such as falls or injuries. Smart houses leverage indoor localization to enhance user experiences, enabling the homeowner to control Wi-Fi network access based on device presence~\cite{Bonafini2019}. Beyond homes, location tracking techniques find applications in agriculture for monitoring greenhouses, libraries for book location, parking garages for car monitoring, and warehouses for item tracking~\cite{MLSurvey-navit}. In disaster management, indoor localization can pinpoint the precise location of individuals in peril for efficient rescue efforts~\cite{Chehri2011,5976190}. Despite these advantages, privacy concerns arise, including the continuous monitoring of sensitive health-related data, potential invasions of privacy in smart homes, perceived intrusive surveillance in various contexts, and privacy implications in disaster management. Striking a balance between the benefits and privacy protection is crucial, necessitating robust data safeguards and ethical guidelines.

\subsection{Health Services and Public Safety} 
Indoor localization holds immense potential for elevating service standards in healthcare, aiding hospital staff in swiftly locating patients and enabling patients to navigate therapy rooms independently. Doctors can monitor patient mobility, visitors can easily find their patients, and operating rooms can be well-equipped~\cite{MLSurvey-navit, CALDERONI2015125}. Smart devices' positions are traceable, facilitating their quick retrieval. The technology's future applications include nanosensor-based drug delivery for tumor targeting and hazard identification for disaster mitigation~\cite{safetymanage}. In law enforcement, Bluetooth beacon-based systems aid quick responses, while in smart buildings, Wi-Fi-enabled alarms guide people to safety. Despite these benefits, concerns arise about continuous tracking compromising patient privacy in healthcare, nanosensor-based drug delivery raising intimate monitoring questions, and surveillance implications in safety management and law enforcement. Striking a balance requires careful ethical and legal considerations in these applications.

During the COVID-19 pandemic, for example, IPSs were employed to monitor and enforce social distancing guidelines in various settings, including hospitals and workplaces. While crucial for public health, these technologies raise significant privacy concerns. The continuous and precise tracking of individuals' movements within indoor spaces may result in the collection of sensitive location data, revealing interactions, routines, and patterns. This raises potential privacy infringements, and there's a risk of unauthorized access or misuse, challenging data security and individual autonomy. Addressing this complex challenge requires robust privacy safeguards, transparent data handling practices, and clear regulations to ensure the responsible deployment of IPS for COVID-19 measures while respecting privacy rights and maintaining public trust. The global COVID-19 pandemic has highlighted the critical importance of privacy in indoor localization systems, especially in contexts where accurate location tracking is necessary to enforce social distancing measures. As outlined in \cite{covid} advancements in privacy-enhancing technologies were key to ensuring that personal data, particularly location information, should be protected while still allowing for real-time monitoring of individuals in shared spaces.

\subsection{Surveillance and Tracking}
LBSs extend beyond human tracking, finding significant applications in industrial settings, where the automatic tracking of numerous objects is essential for effective management. This involves real-time control of item locations alongside localization and identification, requiring new Medium Access Control layer protocols to prevent collisions~\cite{Li2021}. The prediction of autonomous robot locations, especially in the emerging era of collaborative robots (co-bots), enhances safety and efficiency in various tasks, such as safety and collision avoidance in industrial duties~\cite{assettrack,8108561,8371230}. However, the continuous monitoring of objects and devices in industrial contexts raises privacy concerns for workers, impacting workplace privacy. Anticipating robot locations and monitoring employee interactions could similarly affect privacy. Furthermore, the use of location-based access control and continuous tracking for anti-theft systems raises data security and surveillance issues. Additionally, indoor localization in entertainment and broadcasting may involve tracking event attendees and presenting privacy challenges. Achieving a balance between the advantages of these technologies and privacy protection requires the establishment of robust safeguards and guidelines for responsible use~\cite{VERMA2024105041}.

\section{Discussion \& Future Research Directions} \label{sec:Dis}
Indoor location privacy is a multifaceted challenge, given the diverse range of technologies and sensors used in positioning systems. To address these complexities and provide robust privacy protection while enabling the full potential of location-based services, future research can be summarized as follows: 

\subsection{Future Research on Cryptographic Techniques for Indoor Location Privacy}
Based on what is mentioned in Section~\ref{Sec:Cry}, it is clear that the choice of cryptographic technique as an ILFPPM should be carefully aligned with the specific application needs. For real-time systems or applications with limited computational resources, Paillier encryption offers a good balance between privacy protection and performance, making it a suitable choice for scenarios where the adversary model is semi-trusted and the focus is primarily on location privacy. In contrast, for highly sensitive environments with fully malicious adversaries, where both location and data privacy are critical, the use of Garbled Circuits or multi-layer encryption techniques, despite their computational overhead, provides stronger privacy guarantees. It is also recommended to evaluate these cryptographic methods on real-world datasets to ensure practical applicability. Finally, future research should aim to explore hybrid cryptographic methods that can better balance privacy and efficiency, while also addressing protocol-based attacks that exploit weaknesses in communication frameworks.

\subsection{Future Research on Anonymization for Indoor Location Privacy }
Privacy-preserving anonymization techniques should provide a balance between efficiency, scalability, and privacy guarantees. Given the evolving threat landscape, it is crucial to integrate multiple anonymization models (e.g., $k$-anonymity, $\ell$-diversity, $t$-closeness) to address both attribute disclosure risks and location privacy. Additionally, applying these techniques to real-world datasets with varying levels of granularity and adversary models would provide more comprehensive insights into their effectiveness. Future research should also explore hybrid anonymization methods that combine randomization, permutation, and spatial techniques to reduce communication overhead without compromising privacy. Finally, attention should be paid to evaluating anonymization methods under more complex attack models, such as semantic and probability-based attacks, to ensure their resilience in diverse operational environments.

\subsection{Future Research on DP for Indoor Location Privacy}
Future work on DP-based ILFPPM should also aim to refine the balance between privacy protection, data utility, and efficiency. As seen in the evolution from CDP to LDP, decentralized privacy mechanisms show greater potential in protecting sensitive location data in untrusted environments. Researchers should focus on developing hybrid DP methods that combine the strengths of both CDP and LDP to mitigate location and data privacy risks. Additionally, incorporating advanced adversary models, such as semantic and probability-based attacks, into the evaluation of DP methods will provide more comprehensive insights into their resilience. Finally, improving scalability by leveraging edge-cloud collaboration and enhancing run-time efficiency will ensure the practical deployment of these mechanisms in real-world ILF systems.

\subsection{Future Research on FL for Indoor Location Privacy}
Future work should focus on optimizing FL methods for environments with highly variable data distributions, such as multi-building or city-wide systems. Methods like FedProx and federated knowledge distillation show promise but require further exploration to handle non-IID data effectively in large-scale, heterogeneous systems.
While existing works have focused primarily on location privacy and inference attacks, research should explore techniques for defending against emerging attack models like model inversion or MIA. Integrating DP and secure multi-party computation alongside federated techniques could offer more comprehensive defenses.
Moreover, given the importance of efficiency metrics like energy consumption and run time, future research should focus on reducing the computational and communication overhead associated with advanced privacy techniques. Developing lightweight FL models that maintain strong privacy guarantees without compromising performance, particularly for edge devices, will be crucial.
More studies also need to validate FL-based privacy methods using real-world indoor datasets beyond UJIIndoorLoc and private datasets. A wider variety of scenarios and datasets, such as those from smart homes or healthcare facilities, could better demonstrate the effectiveness and generalizability of the proposed methods.


\subsection{Advanced hybrid privacy-preserving techniques} 
As every privacy-preserving technique has its own advantages and limitations, future research endeavors can be towards the development of advanced hybrid ILFPPM. These hybrid techniques aim to overcome the limitations of individual approaches, providing versatile solutions applicable across various sensor modalities and contextual scenarios. By combining different privacy-preserving strategies, these hybrids offer a comprehensive and adaptable framework for robust protection of user data in diverse indoor environments. In response to the evolving attack landscape, advanced hybrid methods can enhance privacy, defending against more potential attacks from breaching user privacy.

\subsection{Context-aware privacy and usability} 
Future research could explore context-aware privacy mechanisms for indoor location data, incorporating user-friendly interfaces. This ensures that users can make informed decisions and manage privacy based on the sensitivity of location data in various contexts. The approach recognizes that privacy preservation is context-dependent and proposes techniques that dynamically adapt to different situations, such as stricter measures in healthcare and more leniency in retail settings. The aim is to develop adaptive privacy solutions capable of addressing concerns across diverse indoor applications.

\subsection{Scalability} 
With the proliferation of IoT devices and the exponential growth of data, scalability challenges will need addressing. Advancements in FL can address scalability challenges by improving model architectures, communication efficiency, optimization algorithms, robustness against various attacks, and aggregation methods in large-scale IPS. 

\subsection{Ethical considerations and privacy standards} 
In response to privacy concerns, many countries have enacted regulations to protect individuals' data. Notable examples include the General Data Protection Regulation (GDPR) in the European Union\footnote{https://gdpr-info.eu/},Health Insurance Portability and Accountability Act (HIPAA)\footnote{https://www.hhs.gov/hipaa/for-professionals/privacy/laws-regulations/index.html} and  the California Consumer Privacy Act (CCPA)\footnote{https://oag.ca.gov/privacy/ccpa} in the United States, and the Personal Information Protection Law (PIPL)\footnote{https://www2.deloitte.com/cn/en/pages/risk/articles/personal-information-protection-law.html} in China. These laws aim to give individuals greater control over their data, including location information. Specifically for location privacy, indoor localization models can ensure GDPR compliance by integrating specialized GDPR-based access control systems~\cite{gdpr-arch}. 

Each of these regulations has distinct requirements for handling location data. The GDPR classifies location data as personal data and mandates explicit user consent, data minimization, and the right to erasure. It also enforces privacy-by-design principles for ILF systems. In contrast, the CCPA grants users the right to know, opt out, and request deletion of their location data but does not require explicit opt-in consent. Meanwhile, PIPL classifies location data as sensitive information, requiring justification for collection, explicit consent, and strict cross-border data transfer regulations.

Given the growing legal study surrounding location data privacy, ILF systems must align with these global regulations. Compliance strategies include: (1) implementing anonymization techniques such as k-anonymity, and DP to prevent re-identification~\cite{FATHALIZADEH2022102665, DP-Ftl2023}, (2) restricting access to location data using cryptographic protection and FL~\cite{survey-2, FL-Liu2019}, (3) minimizing data retention and ensuring compliance with data deletion requests~\cite{gdpr-arch}, and (4) enhancing transparency and user control by allowing users to manage their data and consent preferences~\cite{4343996}. By integrating these legal standards into ILFPPM, service providers can enhance compliance while mitigating the risks associated with location privacy breaches.

\color{black}

\subsection{Privacy benchmarks and collaboration} 
Establishing privacy benchmarks and metrics is crucial for fostering collaboration among academia, industry, policymakers, and advocacy groups in the realm of ILFPPM. This collaborative ecosystem creates a shared framework for evaluating and advancing ILFPPM, incorporating research insights, practical implementations, informed policy perspectives, and user-centric advocacy. Privacy benchmarks act as a crucial link, bridging theoretical expertise with real-world applications to stimulate innovation. Policymakers can leverage these benchmarks to align regulations, ensuring a responsive and effective regulatory framework. User-centric principles, ethical considerations, and a commitment to continuous improvement serve as foundational elements in this collaborative effort, emphasizing responsible innovation and user trust as central tenets in the evolution of indoor location technologies.

\subsection{Multimodal privacy integration} 
In the complex indoor environment, diverse sensor modalities are frequently employed, each possessing distinct strengths and limitations. Advanced approaches offer the capacity to seamlessly integrate data from these various sensors, culminating in comprehension of the indoor area, all while protecting user privacy. For example, the fusion of Wi-Fi and Bluetooth data with inputs from inertial sensors has the potential to significantly elevate not only the precision of indoor positioning but also the efficacy of privacy preservation mechanisms. This integrated approach empowers IPS to harness the complementary strengths of multiple sensor types, thereby enhancing both accuracy and privacy protection, ultimately affording users a more robust and private experience within indoor spaces.

\section{Conclusion} \label{sec:conclusion}
This paper contributes a comprehensive overview of indoor location fingerprinting privacy-preserving mechanisms (ILFPPM), presenting a foundational understanding of privacy preservation in indoor location fingerprinting (ILF) systems.
By addressing various privacy dimensions in such systems, including definitions, applications, vulnerabilities, attacks, metrics, datasets, and protection mechanisms, we aim to provide more insights into the complexities of this dynamic domain. We introduce all possible origins of privacy leakages in ILF systems. The novel categorization of adversary and attack models, along with the compilation of datasets and metrics, serves as a valuable resource for the next empirical investigations. 
Through a detailed exploration of challenges, we point out future research directions and opportunities in the field of privacy-preserving mechanisms for ILF systems. With this extensive survey, we aim to establish a robust groundwork for forthcoming studies in this area.






\bibliographystyle{elsarticle-num}
\bibliography{ref}







\end{document}